\documentstyle[12pt,aasms4]{article}
\def \yskip{\penalty-50\vskip3pt plus 3pt minus 2pt}
\def \pp{\par \yskip \noindent \hangindent .4in \hangafter 1}
\def \abc#1#2#3#4 {\pp#1, {\sl#2}, {\bf#3}, #4}
\def \blank {\lower 5pt\hbox to 0.75in{\hrulefill}}
\def \kms{km~$\rm{s}^{-1}$}
\def \cc{$\rm{cm}^{-3}$}
\def \c2{$\rm{cm}^{-2}$}
\def \lam{$\lambda$}
\def \Ha{H$\alpha$}

\def \mum{$\mu$m}
\def \lsol{L$_{\odot}$}
\def \msol{M$_{\odot}$}

\def \deg{$^\circ$}
\newfont{\rten}{cmr10} 
\def \arcdeg{\hbox{$^\circ$}}
\def \arcmin{\hbox{$^\prime$}}
\def \arcsec{\hbox{$^{\prime\prime}$}}
 
\begin{document}
\normalsize
\title{Infrared and Millimetric Study of the Young Outflow Cepheus E}

\author{Amaya Moro-Mart\'{\i}n}
\affil{Steward Observatory, University of Arizona, Tucson, AZ, 85721}
\affil{Electronic mail: amaya@as.arizona.edu}

\author{Alberto Noriega-Crespo}
\affil{SIRTF Science Center, California Institute of Technology}
\affil{IPAC 100-22, Pasadena, California, 91125}
\affil{Electronic mail: alberto@ipac.caltech.edu}

\author{Sergio Molinari}
\affil{Infrared Processing and Analysis Center,}
\affil{California Institute of Technology}
\affil{Pasadena, California, 91125}
\affil{Electronic mail: molinari@ipac.caltech.edu}

\author{Leonardo Testi}
\affil{Osservatorio Astrofisico di Arcetri}
\affil{50125 Firenze, Italy}
\affil{Electronic mail: lt@arcetri.astro.it}

\author{Jos\'e Cernicharo}
\affil{Instituto de Estructura de la Materia}
\affil{CSIC, Calle Serrano 40, Madrid, Spain}
\affil{Electronic mail: cerni@astro.iem.csic}

\affil{\&} 

\author{Anneila Sargent}
\affil{Division of Physics, Mathematics and Astronomy}
\affil{California Institute of Technology}
\affil{Pasadena, California, 91125}
\affil{Electronic mail: afs@astro.caltech.edu}

\begin{abstract}
The Cepheus E outflow has been studied in the mid and far infrared using 
the ISO CAM and LWS instruments, and at millimetric wavelengths
using OVRO. In the near and mid-IR, its morphology is similar to that
expected for a jet driven outflow, where the leading bow shocks entrain
and accelerate the surrounding molecular gas. As expected, 
fine structure atomic/ionic emission lines arise from the 
bow shocks, at both the Mach Disk and the stagnation tip, 
where J-shocks are dominant. The H$_2$, H$_2$O and CO molecular emission 
could arise further `downstream' at the bow shock wings where the shocks 
(v = $8 - 35$ \kms) are oblique and more likely to be C-type. 
The $^{13}$CO emission arises from entrained molecular gas and a compact 
high velocity emission is observed, together with 
an extended low velocity component that {\it almost} coincides spatially
with the H$_2$ near-IR emission. The millimetric continuum emission shows 
two sources. We identify one of them with IRAS 23011+6126, postulating 
is the driver of the Cepheus E outflow; the other, also an 
embedded source, is likely to be driving one of other outflows observed in 
the region.
Finally, we suggest that the strong [C~II] 158~\mum~emission must originate
from an extended photo-dissociation region, very likely excited by the nearby
Cepheus OB3 association.
\end{abstract}


\section{Introduction}
\label{intro}

Low mass proto-stars produce outflows that 
interact mostly through shocks with the surrounding medium 
(e.~g. Reipurth \& Raga \cite{bo99}). Nevertheless, the details of the 
earliest evolutionary steps remain somewhat unclear.
The embedded flow at Cepheus E (hereafter Cep E) displays
all the characteristics of an extremely young object still surrounded 
by a thick cocoon of cold gas (e.~g. Cernicharo et al. \cite{cerni00}), 
and hence provides an ideal opportunity to study the development of 
molecular/ionic outflows and their surrounding environments during 
the earliest stages of star formation.

Cep E was identified originally as a cloud core in molecular 
CO observations (Sargent~\cite{sar77}), and subsequently, an associated 
outflow was discovered (see Fukui~\cite{fuk89}; Hodapp \cite{hod94}).
The region around Cep  E has proved more complicated than 
the early CO and the K$^{\prime}$ observations indicated;
there are at least two other outflows besides the original compact 
'North-South' H$_2$ flow identified by Ladd \& Hodapp ~(\cite{lad97}).
One is nearly perpendicular to the main H$_2$ flow 
and appears as a faint chain of H$_2$ 1-0 S(1)~2.121~\mum~emission 
knots towards the West (Eisl\"offel et al.~\cite{eis96}); 
the other is seen in the CO J = 1-0 line, is extended by $\sim 90$\arcsec~ 
along a PA $\sim$ -45\arcdeg~(Ladd \& Howe \cite{lad97b}).
All three flows arise within a 5\arcsec~region around IRAS 23011+6126, which
presumably encompasses the driving source 
(Lefloch, Eisl\"offel \& Lazareff \cite{lef96};
Noriega-Crespo, Garnavich \& Molinari \cite{nori98}, hereafter NCGM;
Cernicharo et al. \cite{cerni00}). For a low mass object IRAS 23011+6126 
is estimated to have a very high bolometric luminosity, $\sim~70$~\lsol~
(Lefloch et al.~\cite{lef96}), suggesting that we are dealing with a 
relatively massive protostar, or with a low-mass protostar in an unusually 
early evolutionary phase and considerable higher accretion luminosity.

The blueshifted South lobe of the original H$_2$ Cep E outflow is 
observed at optical wavelengths as Herbig-Haro HH 337, in the shock
excited \Ha\ and [SII] \lam\lam6717/31 emission lines 
(Noriega-Crespo \cite{nori97}; Devine, Reipurth \& Bally \cite{dev97}). 
This collisionally excited emission is characteristic of an extremely 
low excitation HH object ($\sim$~1$\%$ ionization) 
with an anomalously high electron density, n$_e~\sim$~4100~\cc, 
leading to an unusually high pre-shock density 
of $\sim~ 10^5$ \cc ~(Ayala et al. ~\cite{aya00}; hereafter AYA). 
Both the  \Ha\ and [SII] emission can be modeled using a J shock of 
v$_{shock}\sim$~15-30 \kms~(AYA), although
the near-IR H$_2$ spectra agree better with C-shock models with
v$_{shock}\sim$~35 \kms~(Ladd \& Hodapp ~\cite{lad97}). 
It has been suggested that a Mach disk produces the optical emission 
while the IR emission arises from a bow shock (AYA).

Some of the first three dimensional simulations of molecular jets
were intended to model the complex structure of the H$_2$ Cep E 
outflow (Suttner et al. \cite{sut97}). Surprisingly they demanded relatively
large ($>10^5$\cc) densities, which could be a sign of a 
very early stage of jet development. This interpretation is consistent with 
the estimated dynamical age of the Cep E outflow, $\sim~3\times 10^3$ years
(Eisl\"offel et al.~\cite{eis96}), assuming a distance of 730 pc.
The above argument sets a scene in which Cep E is a very young outflow 
breaking through its placental molecular core. 

Our new observations from the {\it Infrared Space Observatory} 
(ISO\footnote{ISO is an ESA project with
the instruments funded by ESA Members States (especially the PI countries: 
France, Germany, the Netherlands and the United Kingdom) and with the 
participation of ISAS and NASA.})
and the {\it Owens Valley Radio Observatory} (OVRO) millimeter wave array
at far/mid infrared and millimetric wavelengths at 
higher sensitivity and spatial resolution, permit not only a probe of 
deeper regions of the molecular environment
surrounding the outflow, but also a determination of the main excitation 
and cooling mechanisms. Both are essential to understand the dynamical 
evolution of very young proto-stellar objects. 

In $\S$ 2  and  $\S$ 3
we describe the ISO and OVRO observations and data reduction procedures,
respectively. The infrared results
are presented in $\S$ 4 and further analyzed in $\S$ 5, where we derive the 
gas physical parameters (such as temperature, density, column density and 
size of the emitting region) and study the origin of this 
emission through J and C-shocks and photo-dissociation regions (PDRs).
In $\S$ 6  we analyze the millimeter results, using the $^{13}$CO 
line emission to derive such physical 
parameters of the outflow as dynamical time scale, mass and mass loss rate,
and continuum observations to provide an alternate estimate of the mass of 
the source.  The source  spectral energy distribution (SED) is examined
in $\S$ 7 and in $\S$ 8 we summarize our results.

\section{ISO Observations and data reduction} 
\label{obs}

The Cep E outflow has been observed with two ISO instruments, 
the Long Wavelength Spectrometer 
(LWS, Clegg et al. \cite{cleg96}) and the infrared camera 
(ISOCAM, Cesarsky et al.~\cite{cesa96}). The observations were taken   
during revolutions 566 (LWS: TDT 56601113 and 56600912) and 792 
(CAM: TDT 79200740).
The LWS01 grating mode was used to acquire low resolution 
(R$\sim$~200) spectra from 43 to 197~\mum ~at two different locations 
along the flow ($\alpha$(2000) = 23h03m13.11s, $\delta$(2000) = 
61\deg42\arcmin59.5\arcsec~ for the North lobe and
$\alpha$(2000) = 23h03m12.73s, $\delta$(2000) = 61\deg41\arcmin56.5\arcsec~ 
for the South lobe). Figure 1 shows the two pointing positions.
These spectra are made up of 11 full grating scans oversampled at 1/4 of a 
resolution element (equivalent to $\sim$~0.07~\mum~for \lam$<$90~\mum~and to
$\sim$~0.15~\mum~for \lam$>$90~\mum~). The integration time per spectral 
element was 22 s, corresponding to a total integration time of 1890 s per 
pointing.
The LWS beam is assumed to be $\sim$~80\arcsec~in diameter 
(ISO Handbook, {\sc iv}, 4.3.2). The ISOCAM (Cesarsky et al.~\cite{cesa96}) 
observations were taken 
using the Circular Variable Filter (CVF) with a total FOV of 
3\arcmin~$\times$3\arcmin, and pixel FOV of 6\arcsec~
centered on IRAS 23011+6126 ($\alpha$(2000) = 23h03m13.02s, 
$\delta$(2000) = 61\deg42\arcmin26.5\arcsec).
The CVF spectra covered the 5.01 $-$ 16.77~\mum~wavelength range with a 
resolving power of $\sim~40$. Each wavelength was observed for $\sim~23$ sec, 
for a total on source observing time of 3550 sec.

The LWS Interactive Analysis \footnote{LIA is available at 
http://www.ipac.caltech.edu/iso/lws/lia/lia.html} (LIA) Package, 
Version 7.2 was used to correct the detector dark currents and drifts. 
Corrected and scan averaged spectra were then analyzed using the ISO Spectral 
Analysis Package\footnote{ISAP is available at 
http://www.ipac.caltech.edu/iso/isap/isap.html} (ISAP) Version 2.0.
Likewise, the CAM-CVF data were reduced using the Cam Interactive Analysis
\footnote{CIA is available at 
http://www.ipac.caltech.edu/iso/cam/camtools.html}
(CIA) standard procedures, and spectra from the individual pixels were 
extracted using XCVF (Ali \cite{ali00}), and further analyzed with ISAP.

\section{Millimeter observations and data reduction}
\label{mili}

High resolution millimeter interferometric observations were obtained 
during 1997 and 2000 using the OVRO millimeter-wave array. 
The array comprises six 10.4 m antennas that were deployed
in three different configurations offering baselines extending 
from the shadowing limit to 220~m. To ensure the best possible sensitivity 
to the extended outflow emission, the data from the 
most compact available configuration is included.  
Each telescope is equipped with cryogenically cooled
SIS receivers providing typical system temperatures of $\sim$~300~K at
110~GHz and $\sim$~1200~K at 222~GHz.  Continuum observations employed an
analog correlator for total bandwidths of $\sim$~2~GHz and 1~GHz at 222 GHz
and 110 GHz (1.3 and 2.7~mm) respectively.  The digital correlator 
was configured to observe simultaneously the $^{13}$CO(1-0) and 
C$^{18}$O(1-0) transitions at 110.2 and 109.8~GHz respectively.  
For the former isotopomer, spectral resolution was 
1.3~\kms~ over a $\sim$~80~\kms~ bandwidth, and for the latter, it was 
0.3~\kms~ over $\sim$~20~\kms.  The half-power beamwidth is 
$\sim$~65\arcsec\ at 110~GHz, and $\sim$~33\arcsec\ at 222~GHz.

Complex gain calibration was ensured by frequent observations of the quasar
BG 0059+58 (= [MFN99] 0059+581). The typical observing cycle alternated scans on Cep E and 
BG 0059+58 with a period of $\sim$~20-30~min. 3C 345 and/or 3C 84 were used as 
passband calibrators while scans on Uranus and/or Neptune were used to set the
flux scale. The expected flux calibration uncertainty is 
$\sim$~15$\%$ at 2.7~mm, and 20-30$\%$ at 1.3~mm.  Data calibration and 
editing have been performed  using the OVRO-MMA software package 
(Scoville et al.~\cite{Sea93}). Imaging and analysis of the
calibrated ($u,v$) datasets were carried out with the NRAO-AIPS package. 
All the maps presented in this paper have been obtained
with the AIPS IMAGR task using natural weighting of the ($u,v$) data for the
molecular outflow maps and the low spatial resolution continuum maps, while
``ROBUST=0'' weighting was used for the 1.3~mm high resolution map.
The beam sizes and noise levels for the OVRO observations are shown in Table 1.

\section{ISO Results}
\label{results}

\subsection{ISO LWS}
Figure 2 shows the LWS far-infrared spectra obtained at the two pointing
positions. The strongest features are the atomic fine 
structure lines of [O~I]~63~\mum~and [C~II]~158~\mum. However,
the spectra are also rich in molecular lines from the rotational 
transitions of CO and H$_2$O (in both ortho and para form). Two OH lines 
are also detected in the spectra, but they are too faint to be included in 
the analysis. 

Integrated line fluxes, measured using a gaussian fit, are shown in Table 2, 
with 1-$\sigma$ uncertainties given in parenthesis. In some cases the FWHM 
was kept fixed to the corresponding instrumental resolution 
element (i.e. 0.29~\mum~ and 0.6~\mum~ for the SW and LW detectors, 
respectively). This technique was particularly important in the case of 
faint or/and blended lines. The uncertainties were obtained either by
using ISAP estimated values from the gaussian fit, or by adopting the 
RMS of the fitted baseline times the resolution element. As expected, both 
values are very similar.  

\subsection{ISOCAM CVF}

CAM CVF provides spectra covering the 5-17~\mum~ range over a 32$\times$32
pixel area.  Each plane of the cube corresponds to a different $\lambda$.
Figure 1 shows, as was previously noticed by NCMG,
how the morphology at 2 and 7~\mum~ in the H$_2$ emission is quite
similar. A more detailed description of the CVF emission at each pixel is 
given in Figure 3. Except at the source position
(pixel (16,16)), the 5 to 16~\mum~ spectra are dominated by the H$_2$ ground
rotational lines from S(7) to S(2). Integrated line fluxes
can be found in Table 3. These are corrected for extinction, using
an estimated E(B-V) = 1.0 (Lefloch et al.~\cite{lef96}; AYA)
and the extinction curve from Draine (\cite{drai89}).
The central source is clearly detected and displays strong silicate absorption 
at 9-12~\mum~and CO$_2$ ice at 15.2~\mum~ and a number of ice bands 
(see also Cernicharo et al. \cite{cerni00}).
Figure 4 displays the images corresponding to the H$_2$ 0-0 S(2) to S(7)  
lines. 
In accord with expectations based on ground based optical and H$_2$ 
observations at 2~\mum, the south lobe is brighter at shorter wavelengths, 
as the flow breaks through the cloud. The north lobe, buried 
deeper in the cloud, has fainter  
S(3)~9.66~\mum~ and S(2)~12.28~\mum~emission, due probably to the broad second 
maxima at 9.7~\mum~ due to silicates (Si-O) stretch mode in the extinction 
curve (Draine~\cite{drai89}, Mathis~\cite{math90}).

\section{Discussion}
\label{discuss}

\subsection{The Gas Physical Parameters}
\label{gas}
\subsubsection{CO and H$_2$O Emission}

The CO and H$_2$O emission can be studied by means of a Large Velocity 
Gradient (LVG) model in spherical geometry, taking into account the first 
35 rotational levels of CO and the first 30 rotational levels of o-H$_2$O.
The collisional rates of Schinke et al. (\cite{sch85}) were used for CO and 
those  of Green, Maluendes and Mc Lean (\cite{gre93}) and Phillips et al. 
(\cite{phi96}) for H$_2$O.

There are six main parameters to fit: temperature (T), H$_2$ 
density (n(H$_2$)), CO and o-H$_2$O column densities (N(CO) and N(o-H$_2$O)), 
intrinsic width of the line (v$_{exp}$) and angular size of the
emitting area ($\Omega$). To determine all of them simultaneously, two
{\it initial} assumptions are necessary: (1) the CO and 
o-H$_2$O emission arise from the same region; (2) the CO lines are optically 
thin, i.e. their line fluxes do not depend
on their intrinsic widths, while the o-H$_2$O lines are 
optically thick.
Since the LWS resolution is too poor to establish size and position
of the emitting region, it is difficult to justify (1) from our data. 
However, in a 15\arcsec~ beam, the H$_2$O
emission in HH 7-11 appears to arise from the shocked regions already seen 
in the CO emission bullets (Cernicharo, Gonzalez-Alfonso \& Bachiller
(\cite{cer96}) and Bachiller \& Cernicharo \cite{bac90}). 
The adoption of assumption (2) will be confirmed by our models.
Figure 5 shows the CO line fluxes as a function of 
the rotational quantum number, J$_{up}$, for 
Cep E North  and South lobes. The shape of the distribution is determined 
only by T and  n(H$_2$), while the absolute fluxes are determined by T, 
n(H$_2$) and $\Omega \times$N(CO), which is proportional to the total number 
of CO molecules within the emitting volume.
  
Two extreme cases for each lobe are consistent with the 
observations.  
All the models predict that the CO lines are 
effectively optically thin. Figure 5 shows that the agreement is good for 
J$_{up}$$<$22; but for higher J$_{up}$ the models underestimate the 
observed fluxes. This could indicate the existence of a hotter CO gas 
component as suggested by Spinoglio et al. (\cite{spin00}) for T Tau.

Assuming that the CO and the o-H$_2$O line emission is  
from the same gas, the temperatures and 
densities derived from the CO lines can be used to model the o-H$_2$O line 
fluxes. Since these lines are optically thick, their absolute fluxes 
depend on all the free parameters in the model, T, n(H$_2$), N(o-H$_2$O), 
v$_{exp}$ and $\Omega$. Their relative fluxes (or line ratios), 
on the other hand, do not depend on $\Omega$. Thus, with T and n(H$_2$)
known, the line ratios were modeled and the values of N(o-H$_2$O) 
and v$_{exp}$ were constrained. Models of the absolute fluxes then 
allowed a derivation of $\Omega$.  
The o-H$_2$O lines at 179, 174, 113 and 75~\mum~ were used because they 
have the highest signal to noise ratio and because the 
o-H$_2$O~179~\mum/174~\mum~ ratio is very sensitive to T and n(H$_2$).

A summary of the LVG model parameters consistent with the observations 
can be found in Table 4. 
A comparison between the LVG models for o-H$_2$O and the observations is
shown in Figure 5. 
The diameter range estimated for the o-H$_2$O emitting region is 
comparable to the extension of the
vibrational H$_2$ 1-0 S(1) 2.12~\mum~emission, 2700-6900 AU 
(3.7\arcsec~ - 9.5\arcsec) at a distance of 730 pc. 
Knowing the emitting area, we
calculate N(CO) and hence the H$_2$O/H$_2$ abundance ratio, assuming a 
standard value of CO/H$_2$$\sim$~10$^{-4}$ in shock regions. 
For H$_2$O, an ortho/para ratio of 3 was adopted, leading to predicted 
p-H$_2$O fluxes within 3-$\sigma$ of the observed values.
The low temperature models give 
H$_2$O/H$_2$$\sim$~4.0$\times$10$^{-5}$ and 1.5$\times$10$^{-5}$ 
for the North and South lobe respectively.
The high temperature models give
H$_2$O/H$_2$$\sim$~2.5$\times$10$^{-4}$ (North) and 
2.4$\times$10$^{-4}$ (South).
Taking into account these estimates together with their uncertainties,
it can be concluded that H$_2$O abundances are between 100 and 4000 
times greater than
expected for the ambient medium in cold quiescent conditions, x(H$_2$O)$\sim$~
10$^{-7}$ (Bergin et al. \cite{berg98}; Moneti, Cernicharo \& Pardo 
\cite{mon00}), 
a result commonly found in young stellar objects 
(Liseau et al. \cite{lise96}; Cernicharo et al. \cite{cer96};
Harwit et al. \cite{harw98}; Gonzalez-Alfonso et al. \cite{gonz98}; 
Ceccarelli et al. \cite{cecca99}; Spinoglio et al. \cite{spin00}). 
This significant enhancement in 
the water abundance within outflow regions can be attributed to the 
vaporization of volatile grain mantles and the increased rate coefficients 
for specific gas-phase reactions at the elevated gas temperatures reached 
behind the shocks, and it is expected to persist for $\sim$~10$^{5}$ years 
(Bergin et al. \cite{berg98}).

According to the LVG models, the high-J CO emission detected by ISO-LWS
indicates a temperature between 220 K and 1200 K. 
Since the contribution to low-J emission from gas at such temperatures is
negligible, the ISO observations must be tracing only the hottest shocked 
molecular material. By contrast, millimetric observations of the 
low-J emission (J=2-1, 
Lefloch et al. \cite{lef96} and J=4-3, Hatchell et al. \cite{hat99}) provide
information about the entrained gas.

\subsubsection{H$_2$ Emission}

The pure rotational H$_2$ lines seen in the ISOCAM CVF spectra arise from 
quadrupole transitions that are always optically thin.  Due to their low 
Einstein coefficients, thermalization is readily possible for moderate volume 
densities.  Hence, LTE calculations are precise
enough to compute H$_2$ column densities. In outflows from 
YSOs, the H$_2$ emission is mostly produced by collisional excitation driven by
shocks, but to gain some insight into the overall
nature of the gas, local thermodynamic equilibrium is commonly assumed.
It is then possible to estimate the excitation temperature ($T_{ex}$)
and  column density of the H$_2$ gas using ln(N$_j$/g$_j$) = ln(N$_{tot}$
/Q) - E$_j$/(k T$_{ex}$), where N$_j$ = 4 $\pi$ F$_j$/($\Omega$ E$_j$ g$_j$ 
A$_j$). As usual, A$_j$ is the transition probability (Turner, Kirby-Docken \& 
Dalgarno~\cite{tur77}) and E$_j$ the energy of the upper level (in cgs units) 
(Dabrowski~\cite{dra84}). The dereddened fluxes, F$_j$/$\Omega$, are listed 
in Table 3. The partition function (Q) was calculated assuming an ortho-para 
ratio of 3.  
The excitation diagrams, ln(N$_j$/g$_j$) {\it vs} 
E$_{up}$ and the derived excitation temperatures and column densities for  
several pixels along the flow are shown in Figure 6. 
Given N$_j$ and $T_{ex}$, a model spectrum can be created for all
H$_2$ 0-0 lines. A comparison one of these models with
the observations is shown in Figure 6. A single
component model predicts the intensities from the S(3) 
to the S(7) lines very well, but the intensity of S(2) is systematically 
underestimated. 

\subsection{Interpreting the Shock Excited Emission}

\subsubsection{C-shocks}

The ground rotational H$_2$ emission observed in YSOs is mostly 
attributed to collisional excitation from shocks, either C-type or J-type 
(Draine \& McKee ~\cite{drai93}). In a nutshell, C-type shocks are 
inherently magnetic and less harsh on molecules because Alfvenic waves 
prepare the gas to the hydrodynamic shock. 
J-type shocks are mostly hydrodynamic and can dissociate H$_2$ 
molecules at lower shock velocities. 
In the mid and far infrared, where molecular transitions are 
detected, C-type shocks are expected to be more important.
This is confirmed by the comparison of the CO and H$_2$ pure rotational 
line fluxes predicted by C-shocks models for v$_{shock}\sim~10-40$
\kms, and gas densities $10^4-10^6$ \cc~(Kaufman \& Neufeld~\cite{kauf96}), 
and J-shocks models for v$_{shock}\sim~30-150$
\kms~and the same gas density range (Hollenbach \& McKee ~\cite{holl89}). 
C-shock fluxes are at least 10 times brighter than the 
J-shocks (and this factor can be much larger
depending on the shock conditions). A similar situation is expected for 
H$_2$O, and the following analysis assumes that only C-shocks
contribute.

The shock velocity and pre-shock density are estimated using the
pure rotational H$_2$ line fluxes and their ratios. From a comparison of these
values with C-shock models (Kaufman \& Neufeld~\cite{kauf96}),
we determine a range of 20$-$40 \kms~shock velocities, but over a wide range 
of pre-shock gas densities. The v$_{shock}$ value of $\sim$~35 \kms~
derived from the near infrared H$_2$ emission 
lines (Ladd \& Hodapp ~\cite{lad97}), falls within this range.
An estimate of the pre-shock gas density can be obtained with the additional
assumption that the emission from the pure rotational H$_2$ and CO lines
arise from the same region. If so, the CO/H$_2$ cooling ratio can be
used in conjunction with the shock velocities.
Since this ratio encompasses all 
cooling, the fraction of lines lying outside the LWS and CVF wavelength 
coverage must be taken into account. The CO, o-H$_2$O and p-H$_2$O cooling
rates derived from the LVG models can be found in Table 4.  For CO, the 
LVG models neglect the contribution 
from lines with J$_{up}$~$>$22, which may belong to a hotter component. 
The H$_2$ cooling was estimated using the
simple LTE model and, after taking into account the difference between the 
LWS and CAM beam, is 1.6$\times$10$^{-18}$~W~cm$^{-2}$ for 
both the North and South lobes. Figure 7 shows the 
comparison between the corrected CO/H$_2$ cooling ratios and
those predicted by C-type shock models (Kaufman \& Neufeld~\cite{kauf96}). 
Within the v$_{shock}$ range of 20$-$40 \kms, the models predict
a pre-shock density of $10^4-2.5\times 10^5$ \cc. 

Figure 8 shows the predicted CO/H$_2$O cooling ratio as a function 
of pre-shock density and shock velocity for C-shocks models 
(Kaufman \& Neufeld~\cite{kauf96}). The observed (and corrected)
ratio for the Cep E outflow indicates shock velocities of 8-14 \kms, 
inconsistent with the values derived from the CO/H$_2$ ratios.  
This could be due to the fact that the brightest emission regions in H$_2$ 
and CO do not seem to coexist spatially, as has been assumed, and therefore
trace different gas conditions. Indeed, $^{13}$CO presented in Figure 9 
suggest this may be the case. We discuss this further below.
The spatial resolution available with LWS and CAM spectra prevent 
further discrimination between shock diagnostics.

\subsubsection{J-shocks and Photo-Dissociation Regions}

The integrated fluxes of the [O~I]~63~\mum~and [C~II]~158~\mum~emission are
comparable for both the North and South outflows (see Table 2). Since the
strength of [C~II] line predicted by shock models is several orders
of magnitude fainter than the [O~I] line, a fraction
of the cooling is due to the presence of a PDR or photodissociation region
(Hollenbach \& Tielens~\cite{holl99}).
Most of the [O~I]~63~\mum~emission is expected to be produced by shock 
excitation, but a percentage comes from a PDR. And so, 
to best use the [O~I]~63~\mum~and [C~II]~158~\mum~lines as diagnostics
of the PDR parameters, an estimate of the [O~I] collisional fraction is
necessary. For this, we use the additional information
provided by the optical forbidden atomic lines 
(e.~g. [O~I] $\lambda$6300, [S~II] $\lambda\lambda$ 6717/31),
and that they are generated by J-shocks, very likely at the Mach Disk (AYA).
Implicitly we are assuming that the molecular emission arises at the
bow shock wings, where C-shocks dominate the emission.

Near infrared and optical observations suggest a range of J-type shock 
velocities of $15-35$~\kms. Assuming 
v$_{shock}$ = 30 \kms, the predicted ratio of [O~I]~63~\mum~to [O~I]~6300 
\AA~ is $\sim~10$ for n = 10$^5$ \cc~ (Hollenbach \& McKee 
~\cite{holl89}). Since the observed [O~I] $\lambda$6300 flux
from the south lobe is $4.7\times 10^{-14}$ erg s$^{-1}$ cm$^{-2}$ 
arcsec$^{-2}$, the predicted [O~I]~63~\mum~flux over the entire object
(r~$\sim~2.5$ \arcsec) is $9.2\times 10^{-12}$ erg s$^{-1}$ cm$^{-2}$. This
is of order $\sim~25\%$ lower than the LWS flux of $12\times 10^{-12}$  
erg s$^{-1}$ cm$^{-2}$. From a comparison of J-shock and C-shock models 
(c.f. Hollenbach \& Mckee ~\cite{holl89}, Draine, Roberge \& 
Dalgarno ~\cite{dra83}; Timmermann ~\cite{tim98}), we estimate that at
n = 10$^5$ \cc~ the contribution of C-shocks to the [O~I]~63~\mum~flux 
is about 5-10$\%$ that of J-shocks.
We will assume that about $\sim~20\%$ of the [O~I]~63 
~\mum~flux  corresponds to the PDR component. However, the percentage
drops as the shock velocity decreases, and indeed for v$_{shock}$ = 20 \kms, 
the collisional emission matches the observed LWS flux.
The corrected ratio of [O~I]~63~\mum~to [C~II]~158 ~\mum~ emission is then 
0.30. With the corrected value of [O~I]~63~\mum/[C~II]~158 ~\mum~, and 
the observed LWS [C~II] flux of 6.8$\times$10$^{-5}$
erg s$^{-1}$ cm$^{-2}$ sr$^{-1}$, PDR models can be used
to determine the far-ultraviolet (FUV) flux and gas densities required to 
reproduce the observations. Two of the models of Kaufman et al.(\cite{kau99})~ 
satisfy the above constrains. For one, the incident FUV flux in units of 
$1.6\times 10^{-3}$erg s$^{-1}$ cm$^{-2}$, G$_0$, is  30, at a density of 
60 \cc, while for the other G$_0$ = 6, at a density of  
5500 \cc. 

The integrated flux in the PAHs bands (e.g. at 6.2, 7.7 
and 11.3~\mum) is correlated with the FUV radiation field (Boulanger et al. 
~\cite{bou98}). Thus the existence of a global FUV field can be indirectly 
inferred from the presence of extended PAHs emission over the entire FOV of 
the CVF  ISOCAM observations (see Figure 3). However, the PAHs emission 
could arise anywhere along the line of sight to Cep E.

Cep E is near  the Cepheus OB3 association (Sargent~\cite{sar77}) with 
members at distances of 500 - 1000 pc (Crawford \& Barnes \cite{cra70}; Jordi,
Trullois \& Galadi-Enr{\'{\i}}quez~\cite{jor96}). Since there are at least
42 bona fide O and B stars in the association (Garmany~\cite{gar73}; Naylor \& 
Fabian~\cite{nay99}), the diffuse
FUV field should be sufficient to generate the observed [C~II] flux. 
One member, HD 217086 (O7n), could by itself account for a FUV flux with 
G$_0\sim$~6 if it is at 725~pc (Garmany~\cite{gar73}), similar to the 
distance to Cep E, and separated from Cep E by $\sim$~80\arcmin. 
Hipparcos measurements of HD 217086 give a parallax of 
1.20$\pm$ 0.92 milli-arcsec, approximately 100 pc farther away, 
but the errors are large.

\section{Results from the Millimetric Observations}
\label{milires}

\subsection{$^{13}$CO Emission}
The Cep E outflow can clearly be seen in the $^{13}$CO J = 1-0 transition at 
110.2 GHz, shown in Figure 9. One of the most interesting features of this 
figure is how the $^{13}$CO emission is almost bounded by the H$_2$ outflow 
lobes, except at the edge of the south flow, which is breaking through the 
molecular cloud. Similar behavior is observed for the high velocity CO gas in 
the HH 211 embedded outflow (Gueth \& Guilloteau~\cite{gg99}; Fig 4),
and in HH 1/2 (Moro-Martin et al.~\cite{amm99}, Fig 2c).

In LTE approximation, using an 
excitation temperature of 20 K (Ladd \& Hodapp \cite{lad97}), 
an abundance of $^{13}$CO/H$_2$ $\sim$~1.2$\times$10$^{-6}$ and 
an inclination angle of 45\deg, the following parameters are derived for the 
red and blue-shifted outflow lobes respectively: dynamical time scales of 
4000 and 8000 years, masses of 0.08 and 0.05 \msol, kinetic luminosities 
of 0.1 \lsol~for both, and mass loss rates of 7$\times$10$^{-6}$ 
and 6$\times$10$^{-6}$ \msol/year. Nevertheless, these estimates are 
quite uncertain due to spatial 
filtering problems, core contamination, and the fact that the edges of the 
outflow are more than 20 \arcsec~ from the phase center of the map, and thus
affected by primary beam attenuation.

Figure 10 shows a position-velocity diagram  along the outflow axis 
(PA=10\arcdeg) with the ambient cloud at -11 \kms and the outflow extending  
approximately from 0 to -20 \kms.
Faint (2-$\sigma$) emission is also seen at $\sim$~30 and -40 \kms. 
We may be tracing high velocity gas that is concentrated nearer to the source 
than its low velocity counterpart. The left panel of Figure 11 shows the 
emission from this gas integrated from -47 to -20 \kms~ and from 
0 to 25 \kms. 
There is no correspondence between this high velocity $^{13}$CO component
and the $^{12}$ CO 4-3 bullets detected by
Hatchell, Fuller \& Ladd (\cite{hat99}). The latter are $\sim~10\arcsec$ away 
from the core and have velocities of -120 and 60 \kms~LSR (a velocity range 
not covered by our observations). A position-velocity diagram along the axis 
of the second outflow detected in $^{12}$CO by Ladd \& Hodapp (\cite{lad97})
(PA=-45\deg) was also constructed but no emission was detected.

Images of the C$^{18}$O J = 1-0 emission at 109.782182 GHz were also made,
and show a core surrounded by a poorly imaged halo. There is no evidence of 
the Cep E outflow, which is not unexpected from this optically thin emission.

\subsection{Continuum Emission at 222 GHz and 110 GHz}

The right panels of Figure 11 display the continuum emission at 
222 GHz and 110 GHz around IRAS 23011+6126. 
The 222 GHz image shows two unresolved 
sources in a 1.38\arcsec$\times$1.06\arcsec~ beam, with integrated fluxes 
of 90 mJy and 67 mJy respectively. At 110 GHz, the continuum emission is 
resolved and has a diameter of $\sim$ 3\arcsec~ or $\sim$~2200 AU (at 730 pc) 
and a flux of 35 mJy. It is tempting to interpret this observations as a 
double source surrounded by an envelope. The presence of a second source 
strongly suggests that the multiple outflows are due to multiple sources 
and not to precession. The projected distance between these sources is of 
$\sim~1.4$\arcsec~or $\sim~1000$ AU at 730 pc. This separation implies a 
long orbital period, even for an intermediate mass system, with a range of
$7\times 10^3$ - $10^4$ yrs, i.e. larger than the kinematical age of 
the outflow.

 Using a dust temperature of 18 K and $\beta$ $\sim$~2 
(Ladd and Howe \cite{lad97b}), a gas-to-dust ratio 
$\sim$~100 and dust opacities of 0.005 cm$^2$g$^{-1}$ 
(Preibisch ~\cite{pre93}), the masses 
associated with the two 222 GHz sources, within 1.4\arcsec, are    
2.5 \msol~and 1.8 \msol. From the 110 GHz continuum, the mass 
associated with the envelope is found to be 13.6 \msol. Ladd \& Howe 
(\cite{lad97b}), 
for comparison, computed a mass of 10 \msol~ within a radius of 3000 AU.  
At 110 GHz, there is no spatial resolution to separate the two sources.

\section{IRAS 23011+6126 Spectral Energy Distribution}
\label{source}

The CVF ISOCAM observations clearly detect an embedded source
that we identified with IRAS 23011+6126. Figure 12 shows the composite 
spectral energy  distribution (SED) of IRAS 23011+6126. It resembles that of 
a Class I source
(Lada \cite{lad87}), rather than a Class 0 as proposed by 
Lefloch et al (\cite{lef96}), based on millimetric observations.
Presently, there are reasons to believe that the distinction 
between Class 0 and Class I is not
as sharp as previously thought (see e.~g. Gregersen et al.~\cite{greg00}).

Following NCGM, model fits to the SED are calculated assuming: (1) a dust 
opacity dominated by  bare silicates at a temperature of 18 K and a density
of 7.5$\times10^4$\cc, and (2) silicates with a thin ice mantle
at a temperature of 18 K and a density of 6$\times10^4$\cc.
The models assume power-law density and temperature distributions, 
with a core inner radius of 0.065 AU and outer radius of 2500 AU.  
As Figure 12 shows, the continuum level of CAM observations near 
log($\nu$) $\sim$~13.3 is lower than that of the bare silicates model.  
As the success of the model using silicates with a thin ice mantle suggests, 
this can be attributed to absorption by H$_2$O at 10-12~\mum~.
The effect has been noticed 
in other low mass protostars such as Elias 29 (Boogert \cite{boo99}).
This model produces a mass envelope of 13.2 \msol~ and bolometric luminosity 
of $\sim~34$ \lsol. Comparable masses have been obtained assuming a constant 
density over a spherical volume 
(10~\msol, Ladd \& Howe \cite{lad97b}) and in dust temperature dependent
models (18~\msol, Lefloch et al.~\cite{lef96}). The bolometric luminosity 
derived by our models, however, is a factor of 2 smaller that these cases 
$\sim~30$~\lsol~ versus 70~\lsol~ or 100~\lsol.
We suspect the discrepancy is due to the integration
methods used, since coarse approximations  overestimate the integrated
flux. Our estimates of the luminosity and envelope mass are more appropriate 
for an intermediate mass object rather than a low mass Class I system; 
although not
as massive and luminous as IRAS 05553+1631 (= GAL 192.16-03.82) (Shepherd et al.~\cite{shep98}) or
IRAS 20126+4104 (Shepherd et al.~\cite{shep00}).

\section{Conclusions} 
\label{con}
The Cep E outflow has been studied in the mid ($\sim 5-17$~\mum) 
and far infrared ($\sim 40-200$~\mum) using the ISO instruments
CAM and LWS, and at millimetric wavelengths (110.2 and 222 GHz) using OVRO. 
In the near and mid-IR, the Cep E morphology is similar to that
expected for a jet driven outflow, where the leading bow shocks entrain and
accelerate the surrounding molecular gas (e.~g. Raga \& Cabrit~\cite{raga93};
Masson \& Chernin~\cite{mass93}). As expected, fine structure 
atomic/ionic emission lines are found, which very likely come from the 
leading bow shock,
in both the Mach Disk and the stagnation tip, where J-shocks are dominant 
(Figure 13). The H$_2$, H$_2$O and CO molecular emission could arise further 
`downstream' at the bow shock wings where the shocks are oblique and more 
likely to be C-type (Smith~\cite{smith91}). The range of shock velocities 
to excite these species is $8-35$ \kms. The $^{13}$CO emission
arises from entrained molecular gas and, as in other outflows, 
a compact high velocity emission is observed, together with an extended low 
velocity component that {\it almost} coincides spatially with the H$_2$ 
near-IR emission.
The millimetric continuum emission shows two sources.
We identify one of them with IRAS 23011+6126 and we postulate 
that it is the driver of the Cep E outflow; the other, also an embedded source,
is likely to be driving one of other outflows observed in the region.
Finally, the strong [C~II] 158~\mum~emission must 
originate from an extended PDR, very likely excited by the nearby Cepheus 
OB3 association.
Our main conclusions are:\nl
\vspace*{0.1cm}
~~1. The mid infrared spectra show strong pure rotational H$_2$ lines, that 
coincide spatially with the H$_2$ 1-0 S(1) 2.12~\mum~emission. 
Excitation diagrams
indicate that this emission is coming from a region at 1000$-$1300 K, in 
agreement with the post-shock temperatures expected for shock velocities
in the range 20$-$30 \kms, consistent with C-shock diagnostics 
of H$_2$ line ratios. \nl
\vspace*{0.1cm}
~~2. The far infrared spectrum is rich in H$_2$O and CO lines.  The line
fluxes can be modeled using an LVG code, assuming that the H$_2$O and CO 
emission arises from the same region.  Two extreme models fit
the observations, with temperatures either of 220 or 1200 K (Table 4). 
The resulting H$_2$O abundances are between 100 and 4000 times 
greater than the one found in quiescent molecular clouds, 
confirming once more
the presence of high water abundance in the shocked gas associated with 
young stellar outflows. From comparison with shock models, we conclude that 
C-shocks with v$_{shock}$ $\sim$~8-14 \kms~ 
can reproduce the H$_2$O and CO observations.\nl
\vspace*{0.1cm}
~~3. Two strong atomic lines [O~I]~63~\mum~
and [C~II]~158~\mum~ are present in the LWS spectra. 
The ratio of their fluxes is close to 1,
indicating a PDR origin.  Taking into account the 
significant contribution to the [O~I] line from J-shocks, and comparing with 
PDR models (Kaufman et al~\cite{kau99}), we find that two models
reproduce the observations, G$_0$ = 30,  n = 60 \cc~and G$_0$ = 6, n = 5500 
\cc~erg s$^{-1}$ cm$^{-2}$, where G$_0$ is the incident FUV flux in 
units of $1.6\times 10^{-3}$ erg s$^{-1}$ cm$^{-2}$.  
Cep E is relatively close to the Cepheus OB3 
association, which can produce the diffuse FUV field required by these 
models.\nl
\vspace*{0.1cm}
~~4. The bipolar outflow can be seen in the high resolution $^{13}$CO
image at 110.2 GHz.  This emission is bounded by the H$_2$ 
condensations. Outflow velocities extend from 0 to -20 \kms, and are 
centered on the ambient
cloud velocity of -11 \kms.  There is also evidence for high velocity gas at 
30 and -40 \kms~ very close to the source.
From the $^{13}$CO observations we estimate,
for the North and South outflows lobes respectively, time 
scales of 4000 and 8000 years, masses of 0.08 and 0.05 \msol, similar kinetic 
luminosities of 0.1\lsol~for both, and mass loss rates of 
$7 \times 10^{-6}$ and $6\times 10^{-6}$ \lsol/year.  
This confirms that the outflow is very 
young as the high water abundance seems to indicate.\nl
\vspace*{0.1cm}
~~5. The continuum images at 222 GHz show the presence
of two unresolved sources within the positional error bars of
IRAS 23011+6126.  The masses associated with them are 2.5 \msol~and 
1.8 \msol. The lower spatial resolution of the continuum 
observations at  110 GHz allow us to detect the surrounding envelope  
which has a mass of 13.6 \msol~ and a radius of 1100 AU.
No evidence is found for any of the other outflows observed
at near-infrared of radio wavelengths.
The presence of a second source, however, strongly suggests that the
multiple outflows are due to multiple sources and not to precession.
The separation of these sources is about 1.4\arcsec or
$\sim~1000$ AU, and so even for an intermediate mass
system, the binary period would be of the order of 
$7\times 10^3-10^4$ yrs, which is longer than the kinematical age 
of the outflow.\nl
\vspace*{0.1cm}
~~6. A simple envelope model for the spectral energy 
distribution of IRAS 23011+6126  predicts a mass envelope of 
13.2 \msol~ and a luminosity of $\sim~30$~\lsol~ from
silicates with thin ice mantles. The spectral features observed at 
5$-$17\mum~in IRAS 23011+6126 are closer to those of a Class I source, 
and the mass and luminosity estimated are those of an intermediate 
mass YSO, rather than a low mass Class 0 object.
\label{end}

\acknowledgements 
We thank Babar Ali of the IPAC ISO group for his insights on the ISOCAM 
data reduction, and John Bieging and the referee for their 
helpful comments. 
A.N-C. \& S.M. work was supported in part by JPL,
California Institute of Technology, under a contract with NASA.
L.T. was partially supported by the Agenzia Spaziale Italiana through
grant ASI ARS-78-1999.
J. Cernicharo thanks Spanish DGES for support under grants
PB96-0883 and PNIE98-1351.
The ISO Spectral
Analysis Package  (ISAP) is a joint development by the LWS and SWS
Instrument Teams and  Data Centers. Contributing institutes are Centre
d'Etude Spatiale des Rayonnements (France), Institute d'Astrophysique
Spatiale (France), Infrared Processing and Analysis Center (United
States), Max-Planck-Insitut f\"ur Extraterrestrische Physisk (Germany),
Rutherford Appleton Laboratories United Kingdom) and the Space Research
Organization, Netherlands.
The Owens Valley millimeter-wave array is supported by NSF grant
AST-99-81546.  Research on young star and disk systems is also supported by
the {\it Norris Planetary Origins Project} and NASA's 
{\it Origins of Solar Systems} program (through grant NAG-5-4591).

{}
 
\makeatletter
\def\jnl@aj{AJ}
\ifx\revtex@jnl\jnl@aj\let\tablebreak=\nl\fi
\makeatother

\begin{deluxetable}{lcc}
\tablewidth{0pc}
\tablecaption{OVRO observations}
\tablehead{
\colhead{Map}  &
\colhead{Beam Size(\arcsec) (PA\deg)}
& \colhead{RMS noise (1-$\sigma$)}
}
\startdata
110 GHz (cont.) & 3.1$\times$2.8 (-86\deg) & 1 mJy/beam \nl
222 GHz (cont.) & 1.4$\times$1.0 (-78\deg) & 7 mJy/beam \nl
C$^{18}$O (1-0) & 8$\times$7 (+63\deg) & 0.05 Jy/beam/channel \nl
$^{13}$CO (1-0) & 8$\times$7 (+54\deg) & 0.03 Jy/beam/channel \nl
\enddata
\end{deluxetable}

\begin{deluxetable}{lcccc}
\tablewidth{0pc}
\tablecaption{Cep E LWS Integrated Fluxes\tablenotemark{a}}
\tablehead{
\colhead{$\lambda$($\mu$m)} & \colhead{Element}  &
\colhead{Transition} & \colhead{North} & \colhead{South}
} 
\startdata
63.18  &  [O~I]     & 3P$_1$-3P$_2$     &  9.3(0.5)    & 12.1(0.7)\nl
75.38  &  o-H$_2$O & 3$_{21}$-2$_{12}$ & 1.8(0.4)       & 0.9(0.2)\tablenotemark{b} \nl
78.74  &  o-H$_2$O & 4$_{23}$-3$_{12}$ & 0.6(0.5)\tablenotemark{b}       & \nodata \nl
79.36 &   CO       & 33-32             & 1.3(0.4)\tablenotemark{b}       & \nodata \nl
82.03  &  o-H$_2$O & 6$_{16}$-5$_5$  & \nodata      & 0.9(0.4)\tablenotemark{b}  \nl
84.60  &  OH       & 3/2-3/2 (7/2--5/2+ 4-3) & 1.5(0.4)\tablenotemark{b}      & 0.8(0.3)\tablenotemark{b}  \nl
89.99  &  p-H$_2$O & 3$_{22}$-2$_{11}$ & \nodata & 1.0(0.2)\tablenotemark{c} \nl
96.77  &  CO       & 27-26             & 1.2(0.2)       & 0.8(0.1) \nl
99.49  &  o-H$_2$O & 5$_{5}$-4$_{14}$  & 1.2(0.4)\tablenotemark{d}     & 0.7(0.3)\tablenotemark{d} \nl
100.46 &  CO       & 26-25             & 0.6(0.4)\tablenotemark{d}   & 0.8(0.6)\tablenotemark{d} \nl         
100.91 &  o-H$_2$O & 5$_{14}$-4$_{23}$ & 1.0(0.4)\tablenotemark{d}      & \nodata \nl
104.44 &  CO       & 25-24             & 0.7(0.3)       & \nodata \nl
108.07 &  o-H$_2$O & 2$_{21}$-1$_{10}$ & 2.6(0.5)\tablenotemark{d}     & 0.9(0.4)\tablenotemark{d} \nl
108.76 &  CO       & 24-23             & 1.0(0.8)       & 1.1(0.6) \nl
113.54 &  o-H$_2$O & 4$_{14}$-3$_{3}$  & 2.7(0.3)\tablenotemark{f}     & 1.6(0.3)\tablenotemark{d,f} \nl
119.44 &  OH       &3/2-3/2 (5/2+-3/2- 2-1) & 0.3(0.1) & 0.5(0.1)\nl
125.35 & p-H$_2$O  & 4$_{4}$-3$_{13}$  & 0.3(0.1)     & 0.4(0.2) \nl
126.71 & p-H$_2$O  & 3$_{31}$-3$_{22}$ & 0.4(0.2)    & 0.3(0.2) \nl
130.37 &  CO       & 20-19             & 0.7(0.1)     & 0.8(0.1)\tablenotemark{d} \nl
137.20 &  CO       & 19-18             & 0.7(0.1)     & 1.0(0.2)\tablenotemark{d}\nl
138.53 &  p-H$_2$O & 3$_{13}$-2$_{2}$  & 0.3(0.1)     & 0.5(0.2)\tablenotemark{d} \nl
144.78 &  CO       & 18-17             & 1.2(0.2)\tablenotemark{d}     & 1.4(0.4)\tablenotemark{d} \nl 
145.52 &  [O~I]     & 3P$_0$-3P$_1$     & $<$0.3\tablenotemark{d}       & 0.9(0.6)\tablenotemark{d} \nl
153.27 &  CO       & 17-16             & 1.0(0.1)     & 1.9(0.1)\tablenotemark{d} \nl
157.74 &  [C~II]    & 2P$_{3/2}$-2P$_{1/2}$ & 7.0(0.3)\tablenotemark{d} & 8.1(0.2)\tablenotemark{d} \nl
162.81 &  CO       & 16-15             & 1.7(0.1)     & 1.9(0.1)\tablenotemark{d}  \nl
173.63 &  CO       & 15-14             & 2.1(0.6)\tablenotemark{d}     & 2.2(0.6)\tablenotemark{d}  \nl
174.62 &  o-H$_2$O & 3$_{3}$-2$_{12}$  & 1.6(0.9)\tablenotemark{d}     & 1.5(1.0)\tablenotemark{d} \nl
179.53 &  o-H$_2$O & 2$_{12}$-1$_{1}$  & 2.7(0.2)\tablenotemark{d}     & 3.1(0.4)\tablenotemark{d} \nl
185.99 &  CO       & 14-13             & 2.1(0.2)     & 2.3(0.2)\tablenotemark{d}  \nl
\tablenotetext{a}{In units of 10$^{-19}$W/cm$^{2}$. 1-$\sigma$ uncertainties are given in parenthesis.}
\tablenotetext{b}{FWHM was fixed at 0.29$\mu$m.}
\tablenotetext{c}{Blended with CO 29-28 90.16$\mu$m.}
\tablenotetext{d}{FWHM was fixed at 0.60$\mu$m.}
\tablenotetext{e}{Blended with p-H$_2$O  2$_{20}$-1$_{11}$ 100.98$\mu$m.}
\tablenotetext{f}{Blended with CO 23-22 113.76$\mu$m.}
\tablenotetext{}{Horizontal dotted lines indicate non detections}
\enddata
\end{deluxetable}

\clearpage

\begin{deluxetable}{lccccccc}
\tablewidth{0pc}
\tablecaption{Rotational H$_2$ 0-0 Observed Fluxes\tablenotemark{a}}
\tablehead{
\colhead{Line($\mu$m)}  &
\colhead{(15,12)} & \colhead{(15,13)} & \colhead{(15,14)} & \colhead{(17,17)} & \colhead{(17,18)} & \colhead{(17,19)} & \colhead{(19,21)} } 
\startdata
S(7) 5.51  & 3.9(0.6) & 4.3(0.5) & 2.0(0.3) & 2.3(0.3) & 7.5(0.5) & 4.1(0.2) & 1.5(0.3) \nl
S(6) 6.11  & 3.1(0.5) & 3.3(0.4) & 1.0(0.2) & 0.9(0.2) & 4.1(0.2) & 2.2(0.3) & 1.5(0.3) \nl
S(5) 6.91  & 5.7(0.2) & 7.2(0.3) & 3.8(0.2) & 5.1(0.3) & 12.7(0.6)& 6.5(0.3) & 3.3(0.1) \nl
S(4) 8.02  & 3.5(0.2) & 4.2(0.2) & 2.0(0.2) & 1.4(0.2) & 2.6(0.2) & 1.8(0.1) & 1.6(0.1) \nl
S(3) 9.66  & 2.5(0.2) & 4.3(0.2) & 3.2(0.1) & 1.9(0.1) & 5.2(0.3) & 3.8(0.2) & 1.9(0.1) \nl
S(2) 12.28 & 2.0(0.2) & 4.0(0.3) & 3.1(0.2) & 1.4(0.1) & 1.9(0.2) & 1.6(0.2) & 1.3(0.2) \nl
\tablenotetext{a}{Fluxes dereddened by E(B-V) = 1.0, in units of 10$^{-11}$~W/cm$^2$/sr}
\enddata
\end{deluxetable}

\clearpage
\makeatletter
\def\jnl@aj{AJ}
\ifx\revtex@jnl\jnl@aj\let\tablebreak=\nl\fi
\makeatother

\begin{deluxetable}{lcc}
\tablewidth{0pc}
\tablecaption{LVG parameters}
\tablehead{
\colhead{}  &
\colhead{~~North} & \colhead{South}
}
\startdata
Temperature (K)                 & ~220~~~~~~1200 & 215~~~~~1200\nl
n(H$_2$) (10$^6$cm$^{-3}$)      & ~2.0~~~~~~~0.02 & 4.0~~~~~0.04\nl
v$_{exp}$ (\kms)                & ~~20-40~~~~~25-40& 30-40~~~~15-60\nl
Source Diameter ('')            & ~~~5.1-3.7~~~5.7-4.3& ~5.8-5.1~~~9.5-4.7\nl
N$_{CO}$ (10$^{17}$cm$^{-2}$)   & ~~9.5-18.0~~~16-26& ~~7.3-9.3~~~2.8-11.3\nl
N$_{o-H2O}$ (10$^{17}$cm$^{-2}$)& ~~3-5~~~~~~~30-50& 0.8-1.0~~~~5-20\nl
N$_{p-H2O}$ (10$^{17}$cm$^{-2}$)& ~~~1-1.7~~~~~10-16.7& ~0.27-0.3~~~1.7-6.7\nl
L$_{CO}$ (10$^{-18}$ W cm$^{-2}$)\tablenotemark{a}&~2.1~~~~~~~~2.6 & 2.4~~~~~~2.8\nl
L$_{o-H2O}$ (10$^{-18}$ W cm$^{-2}$)\tablenotemark{b}&~1.8~~~~~~~~2.1 & 1.3~~~~~~2.1\nl
L$_{p-H2O}$ (10$^{-18}$ W cm$^{-2}$)\tablenotemark{c}&~0.6~~~~~~~~0.6 & 0.4~~~~~~0.6\nl
\tablenotetext{a}{Observed CO fluxes are 14.3(1.2)$\times$ 10$^{-19}$ (North) and 14.2(1.2)$\times$ 10$^{-19}$ W cm$^{-2}$ (South)}
\tablenotetext{b}{Observed o-H$_2$0 fluxes are 14.2(1.4)$\times$ 10$^{-19}$ (North) and 9.6(1.3)$\times$ 10$^{-19}$ W cm$^{-2}$ (South)}
\tablenotetext{c}{Observed p-H$_2$0 fluxes are 1(0.3)$\times$ 10$^{-19}$ (North) and 2.2(0.4)$\times$ 10$^{-19}$ W cm$^{-2}$ (South)}

\enddata
\end{deluxetable}

\clearpage

\begin{center}
Figure Captions
\end{center}

\figcaption[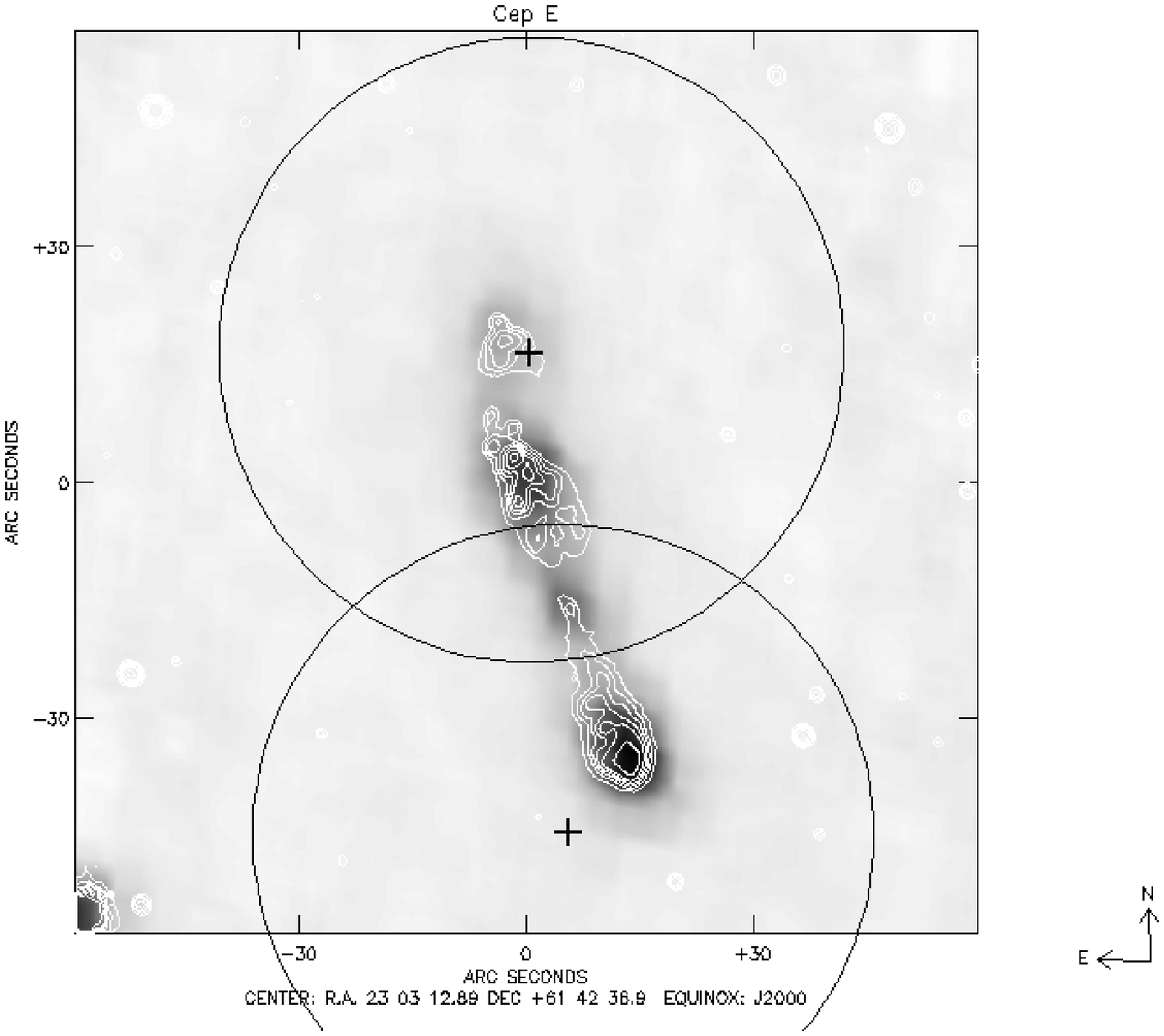]{ISOCAM-CVF observations of the Cep E outflow 
at H$_2$ 0-0 S(5) 6.1$\mu$m. The grey scale depicts the H$_2$ 0-0 S(5)~6.91
~\mum~continuum-subtracted emission, and the superimposed contour image is the NIR vibrational
H$_2$~ 1-0 S(1) 2.121~\mum~line (AYA). The FOV is $\sim$~2\arcmin.
The locations of the LWS circular apertures  are also shown in the image
(r$\sim~40$\arcsec), where the crosses represent the two pointing positions. 
\label{f1}} 

\figcaption[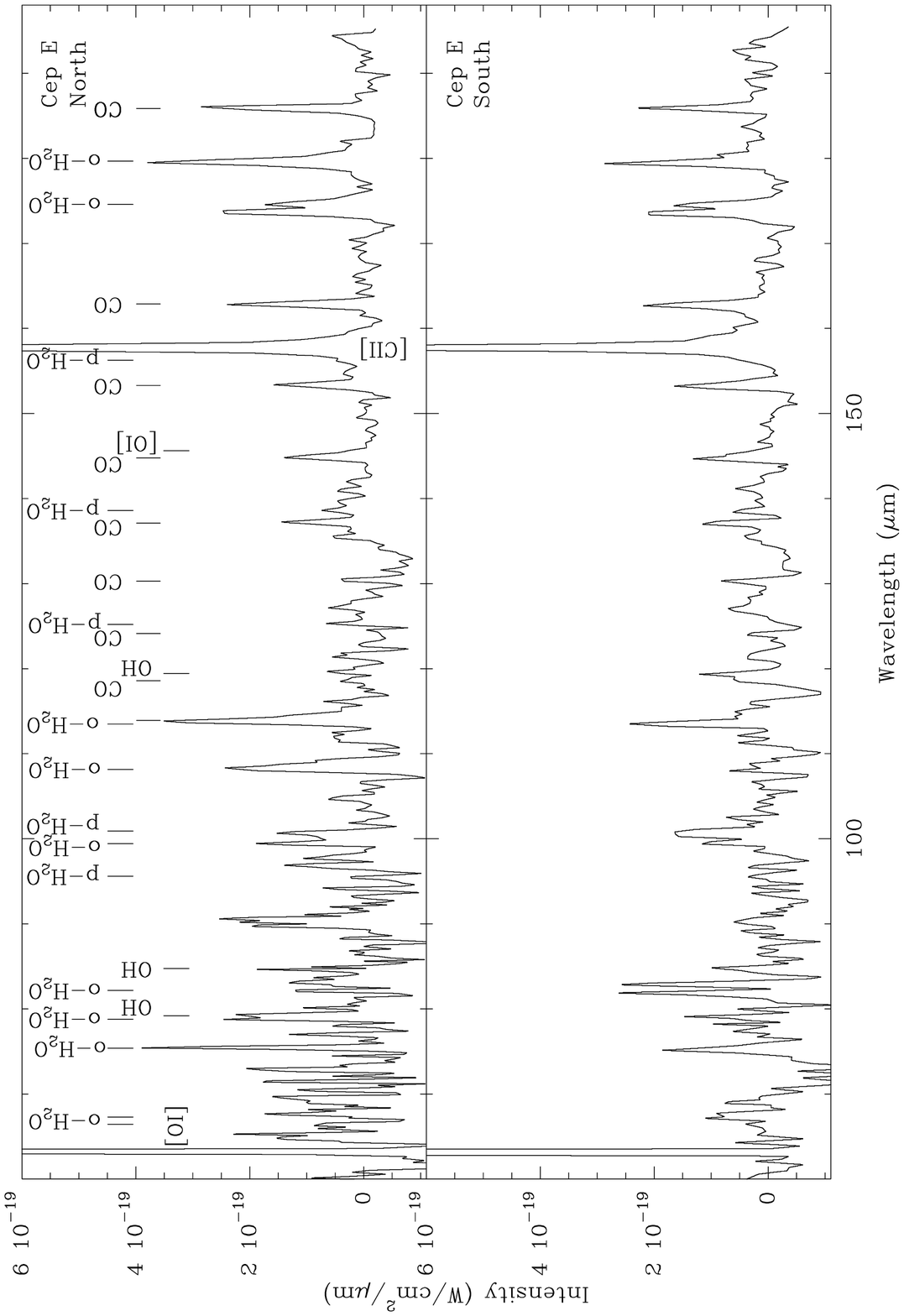]{ISO-LWS spectra of the Cep E North and South lobes
after continuum subtraction and averaging over all detectors. Integrated line 
fluxes are given in Table 2.
\label{f2}} 

\figcaption[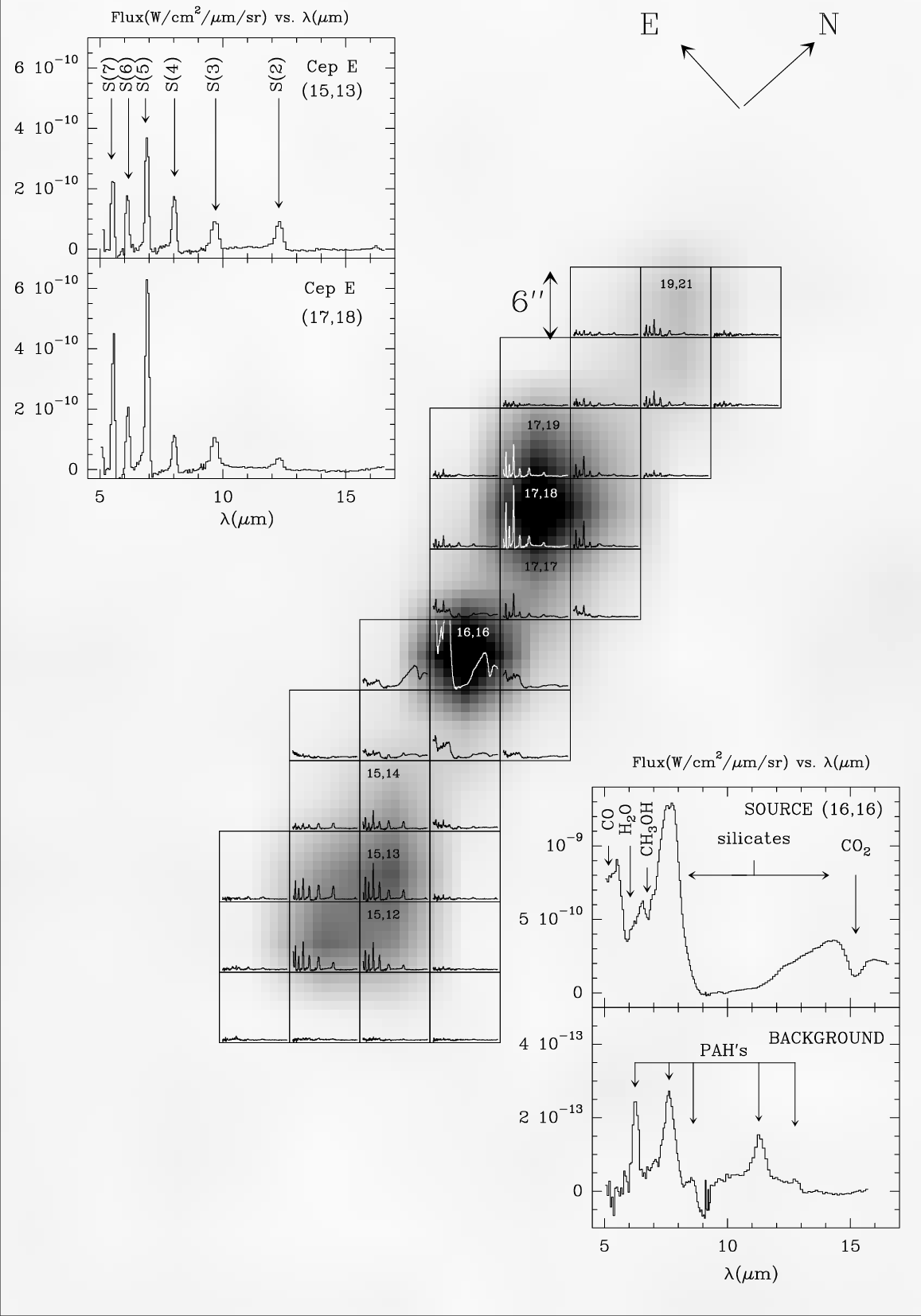]{ISOCAM-CVF observations of Cep E outflow. In 
grey scale is the H$_2$ 0-0 S(5) 6.91$\mu$m image. Superimposed are the 
background subtracted spectra obtained at several positions along the 
outflow. Their scale is identical to that of the 
two magnified spectra shown in the upper left, for one position in the South  
and one in the North lobe.  The spectrum at the central source position,
IRAS 23011+6126, is shown in the lower right on a different scale. Below it, 
is the spectrum of the background averaged over 100 
pixels. Orientation and pixel size are indicated in the figure.
\label{f3}}

\figcaption[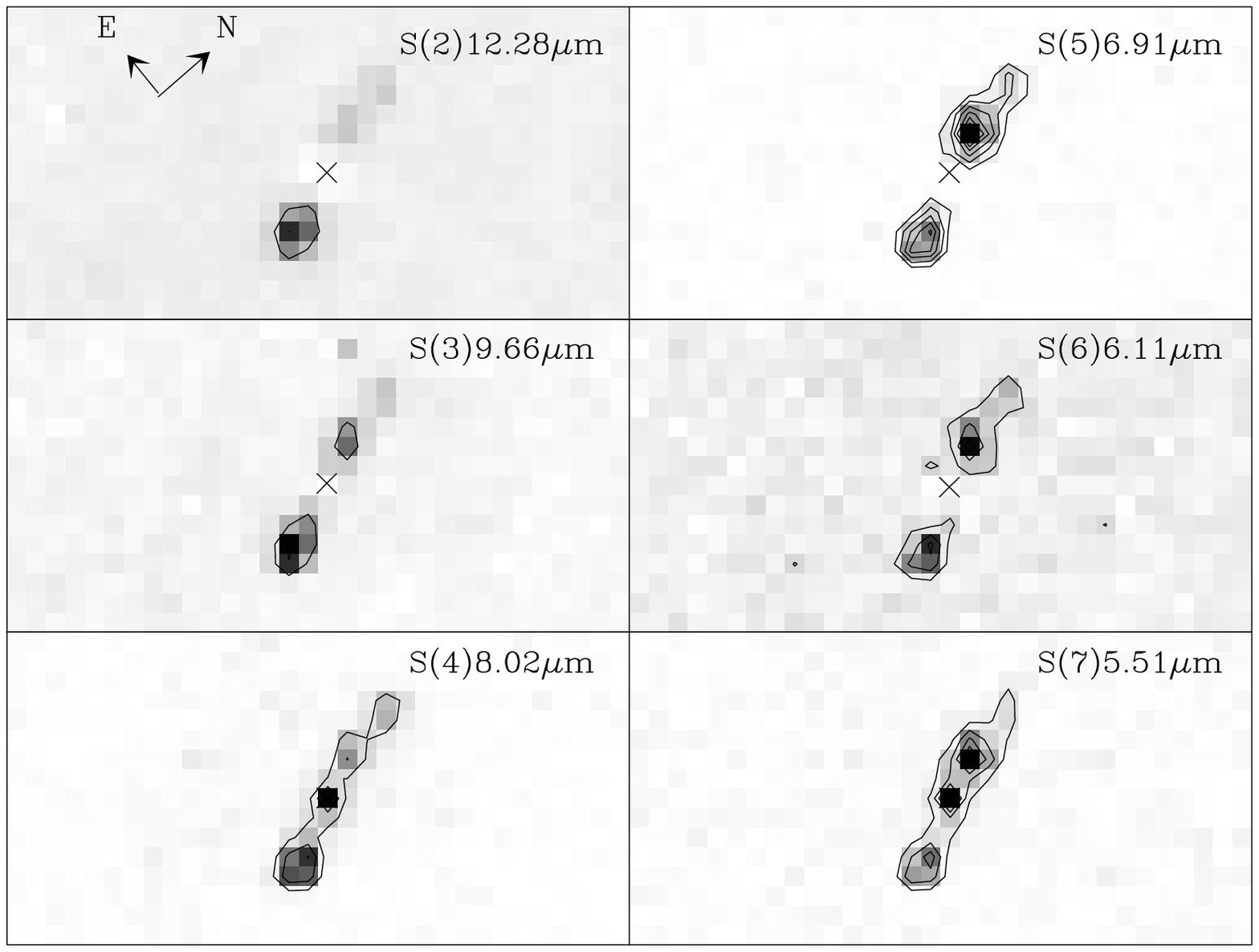]{ISOCAM CVF images of the Cep E outflow at the 
rotational H$_2$ 0-0 lines, after the emission from the adjent line-free 
planes have been subtracted.
All contour levels are from 0.3 to 3.1 by 0.4, in units of 
$10^{-9}$ W/cm$^2$/sr. The figures have been plotted in different 
grey scales to enhance the contrast. The crosses indicate the position of the
source.
\label{f4}} 

\figcaption[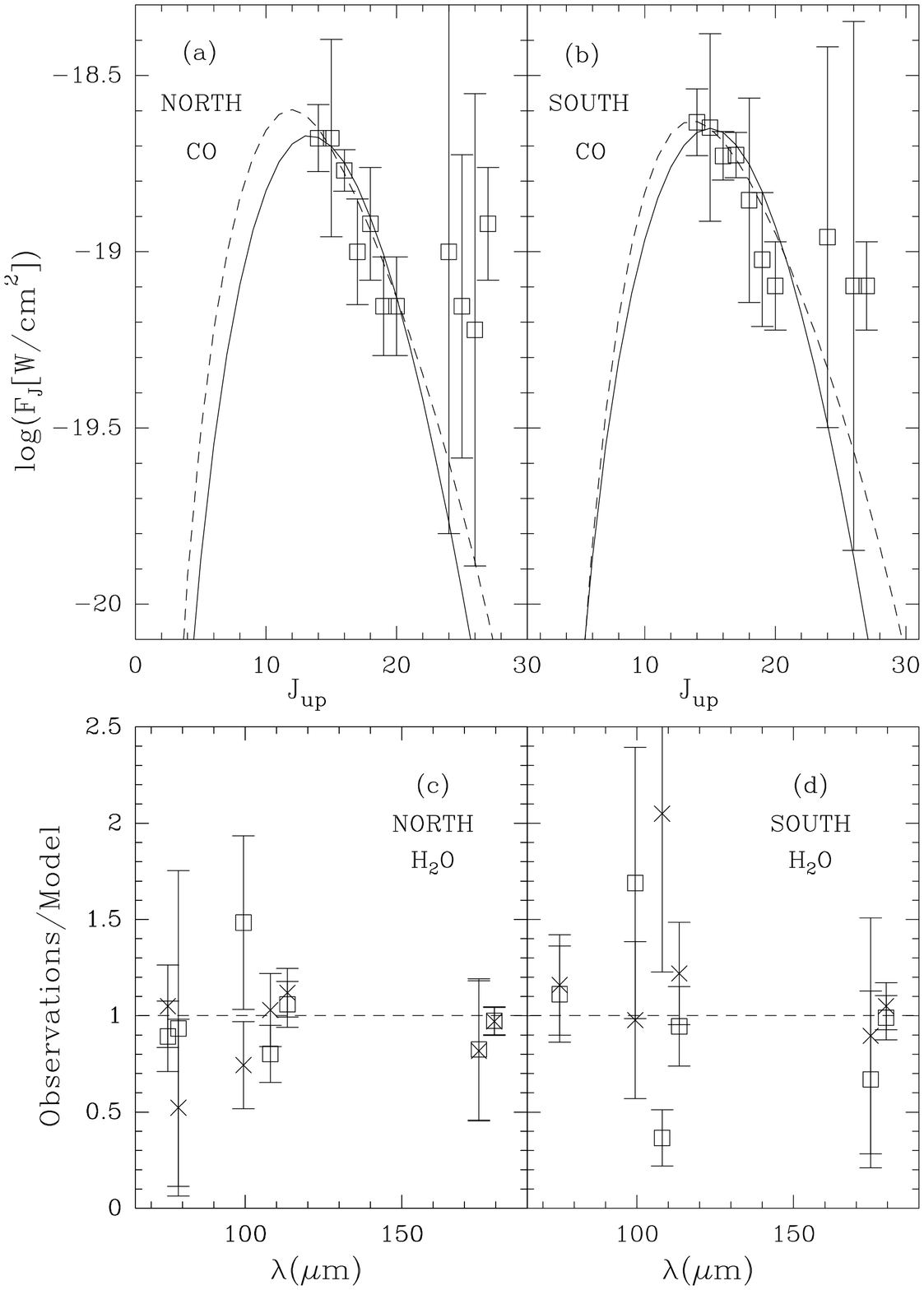]{Comparison between the LVG predictions and 
the observations for CO and H$_2$O lines fluxes. 
(a)~CO line fluxes versus the 
rotational quantum number J$_{up}$ for Cep E North lobe.  Error bars indicate 
1-$\sigma$ uncertainties. The two extreme LVG models that fits the 
observations are shown in solid 
(T = 220 K, n(H$_2$) = 2$\times 10^6$ cm$^{-3}$,
N(CO) = 8.65$\times 10^{17}$ cm$^{-2}$, D = 5.1\arcsec, v$_{exp}$ = 20\kms) 
and dashed 
(T = 1200 K, n(H$_2$) = 2$\times 10^4$ cm$^{-3}$, 
N(CO) = 1.6$\times 10^{18}$ cm$^{-2}$, 
D = 5.7\arcsec, v$_{exp}$ = 25 \kms) lines. 
(b)~Same as (a) for Cep E South lobe. 
The two LVG models are in solid (T = 215 K, 
n(H$_2$) = 4$\times 10^6$ cm$^{-3}$, 
N(CO) = 7.3$\times 10^{17}$ cm$^{-2}$, 
D = 5.8\arcsec, v$_{exp}$ = 30 \kms) and dashed (T = 1200 K, 
n(H$_2$) = 4$\times 10^4$ 
cm$^{-3}$, N(CO) = 1.1$\times 10^{17}$ cm$^{-2}$, D = 15\arcsec, 
v$_{exp}$ = 50 \kms) lines. 
(c) Ratio between LWS fluxes of o-H$_2$O lines and 
LVG model predictions for Cep E North lobe.  Error bars represent 1-$\sigma$ 
uncertainties. Open squares correspond to the low temperature LVG model 
(same parameters as above with 
N(o-H$_2$O) = 3$\times 10^{17}$ cm$^{-2}$), while 
crosses correspond to the high temperature model 
(with N(o-H$_2$O) = 3$\times 10^{18}$ cm$^{-2}$).  
(d) Same as (c) for Cep E South 
lobe. Open squares correspond to the low temperature LVG model 
(with N(o-H$_2$O) = 8$\times 10^{16}$ cm$^{-2}$), while 
crosses correspond to the high temperature model 
(with N(o-H$_2$O) = 1$\times 10^{17}$ cm$^{-2}$)
\label{f5}} 

\figcaption[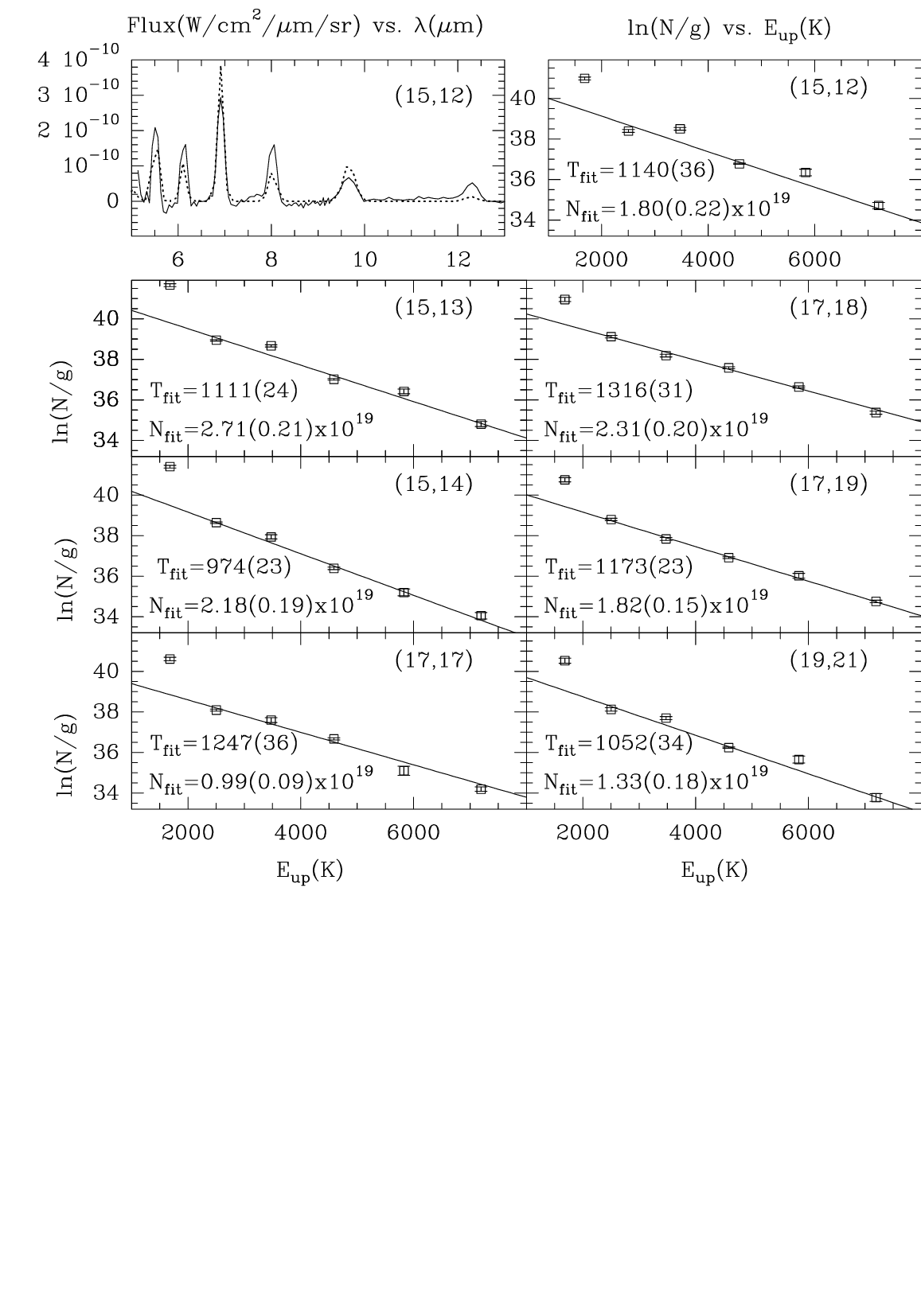]{Determination of the H$_2$ temperature and 
column density from the H$_2$ 0-0 rotational lines observed with ISOCAM-CVF 
(corrected by extinction, E(B-V)=1.0). 
The first panel is the dereddened observed spectra(solid line) at 
pixel (15,12), together with the model (dotted line) derived from the
fitted temperature and column densities. On the right
is the excitation diagram and the best-fit parameters for this pixel.
The rest of the panels show the excitation diagrams for six
positions along the flow with their corresponding best-fit values.
\label{f6}} 

\figcaption[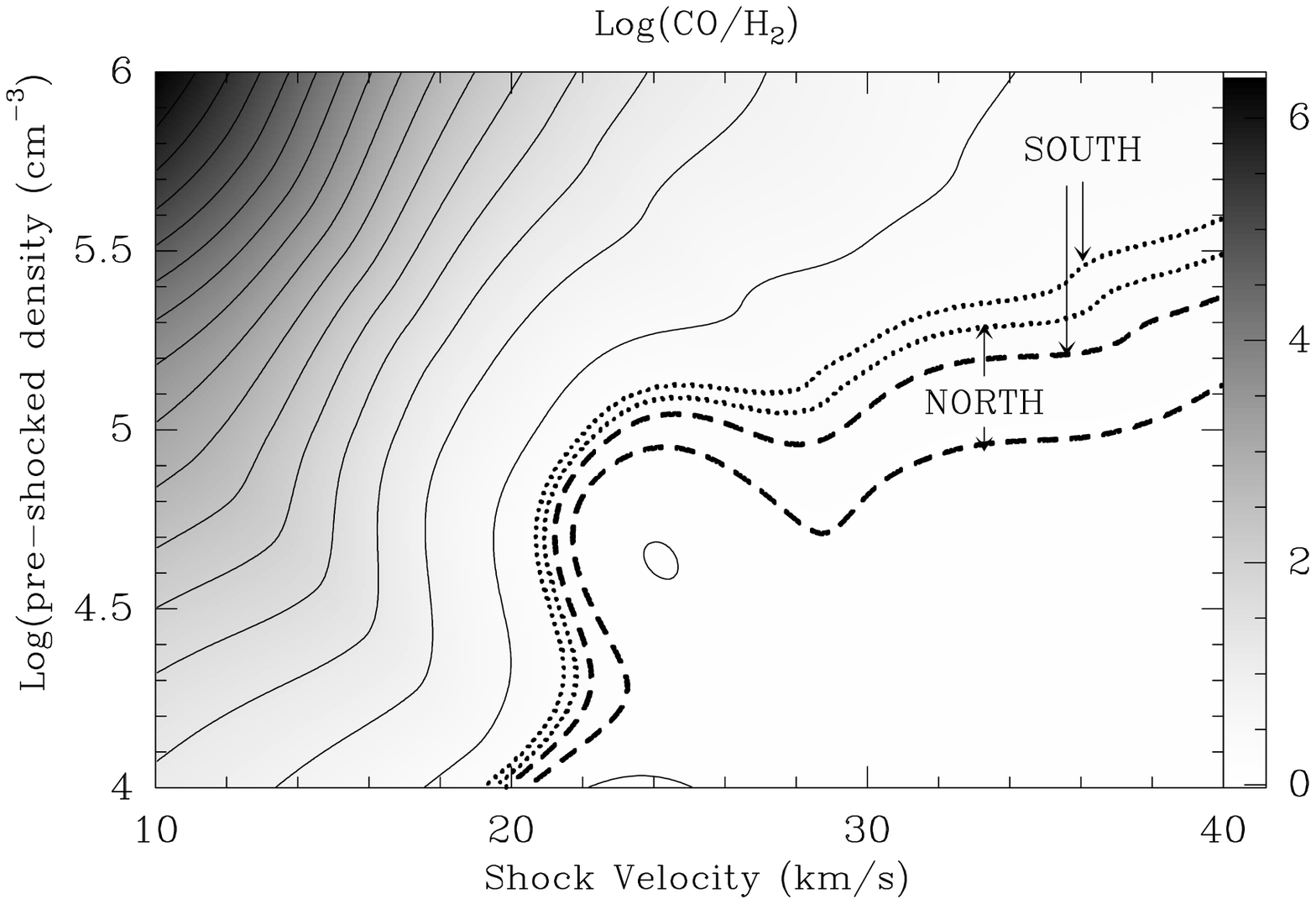]{CO/H$_2$ (rotational) cooling ratio (grey
scale) as a function of pre-shock density and shock velocity for C-shocks 
models (with B = (n$_H$/cm$^{-3}$)$^{1/2}$ $\mu$G; from Kaufman and 
Neufeld 1996). The coolings obtained from the 200 K
and 1200 K LVG models for Cep E North and South lobes are shown in dashed and 
dotted lines respectively. Contour levels are from 0 to 6.4 by 0.4.
\label{f7}} 

\figcaption[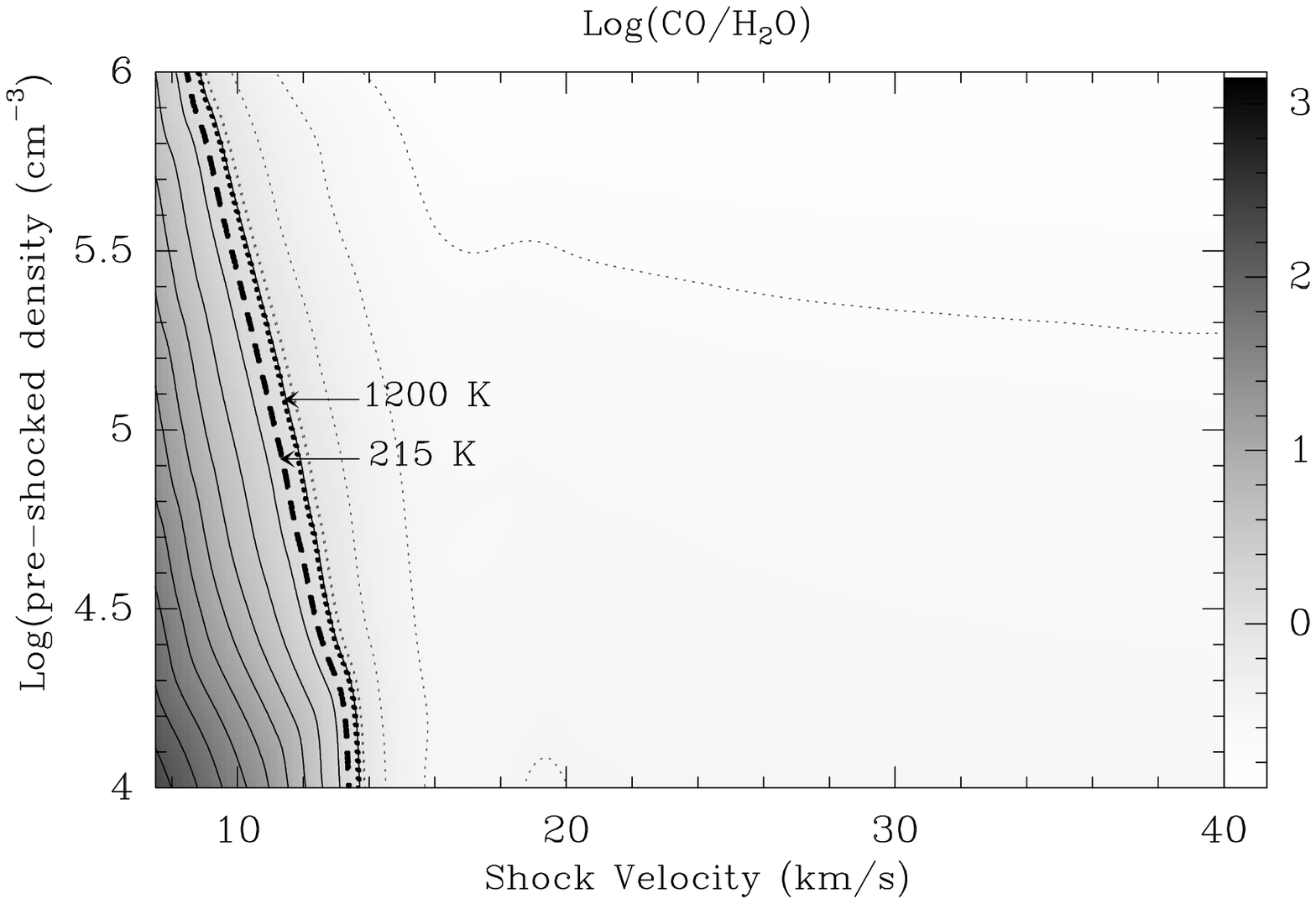]{CO/H$_2$O cooling ratio (grey scale) as a 
function of pre-shock density and shock velocity for C-shocks models
(with B = (n$_H$/cm$^{-3}$)$^{1/2}$ $\mu$G; from Kaufman and Neufeld 1996). 
The coolings obtained from the 200 K
and 1200 K LVG models for Cep E North and South lobes are shown in dashed and 
dotted lines respectively. Contour levels are from -1 to 2.25 by 0.25.
\label{f8}} 

\figcaption[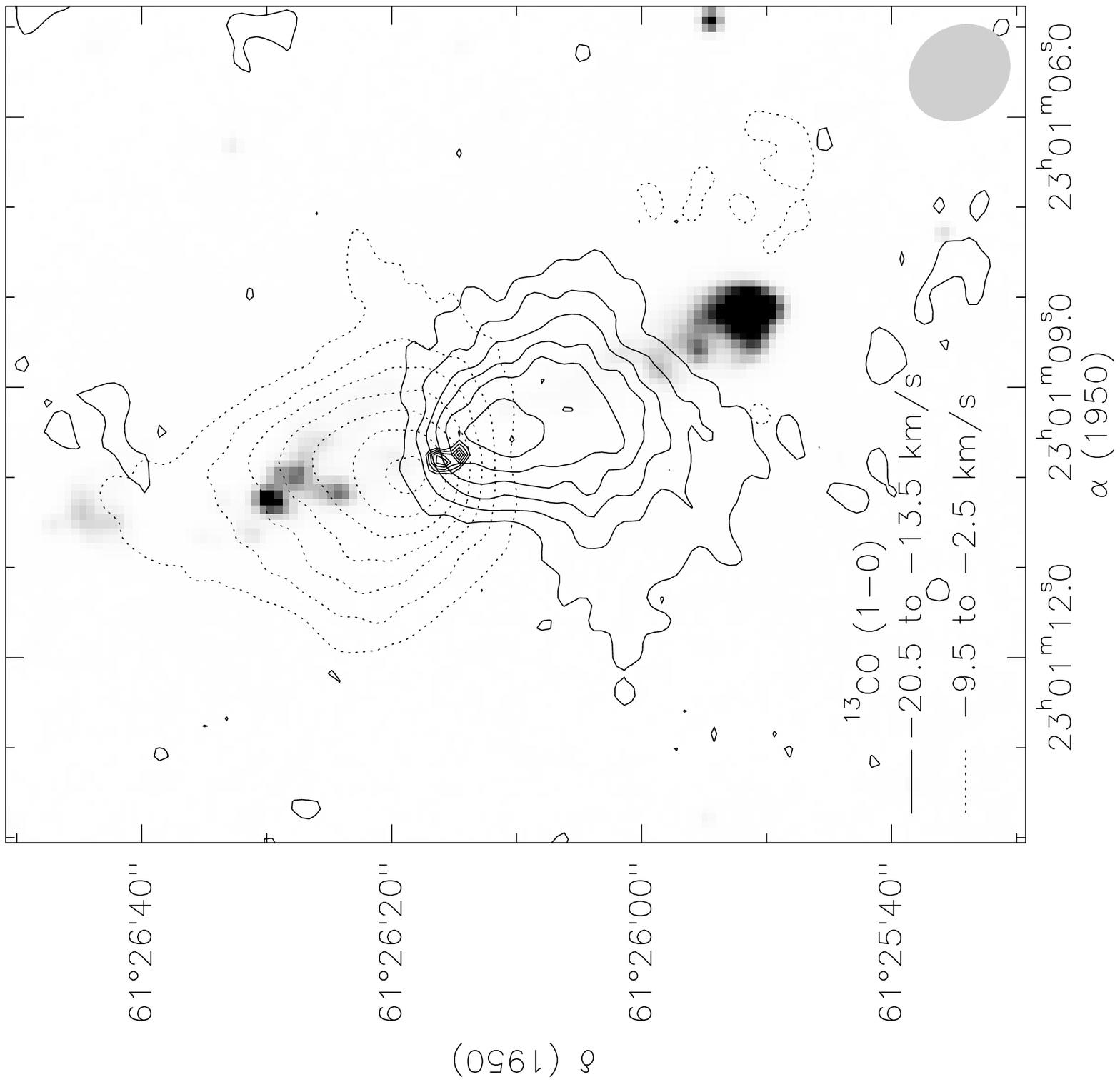]{
Superposition of the H$_2$~ 1-0 S(1) 2.12~\mum~emission 
(in grey scale) on the $^{13}$CO and 222 GHz continuum in contours.  
The dashed contours towards the north
correspond to the red-shifted lobe (v$_{LSR}$ = -9.5 to -2.5 \kms), 
while the solid contours to the south correspond to the blue-shifted lobe 
( v$_{LSR}$ = -20.5 to -13.5 \kms).
Contour levels for $^{13}$CO are 0.4 to 3.0 by 0.4 and 0.2 to 1.4 by 0.2 
respectively (in Jy/beam). Contour levels for 222 GHz are from 0.02 to 
0.05 by 0.01 (in Jy/beam). The beam size is showed at the lower right corner.
\label{f9}}

\figcaption[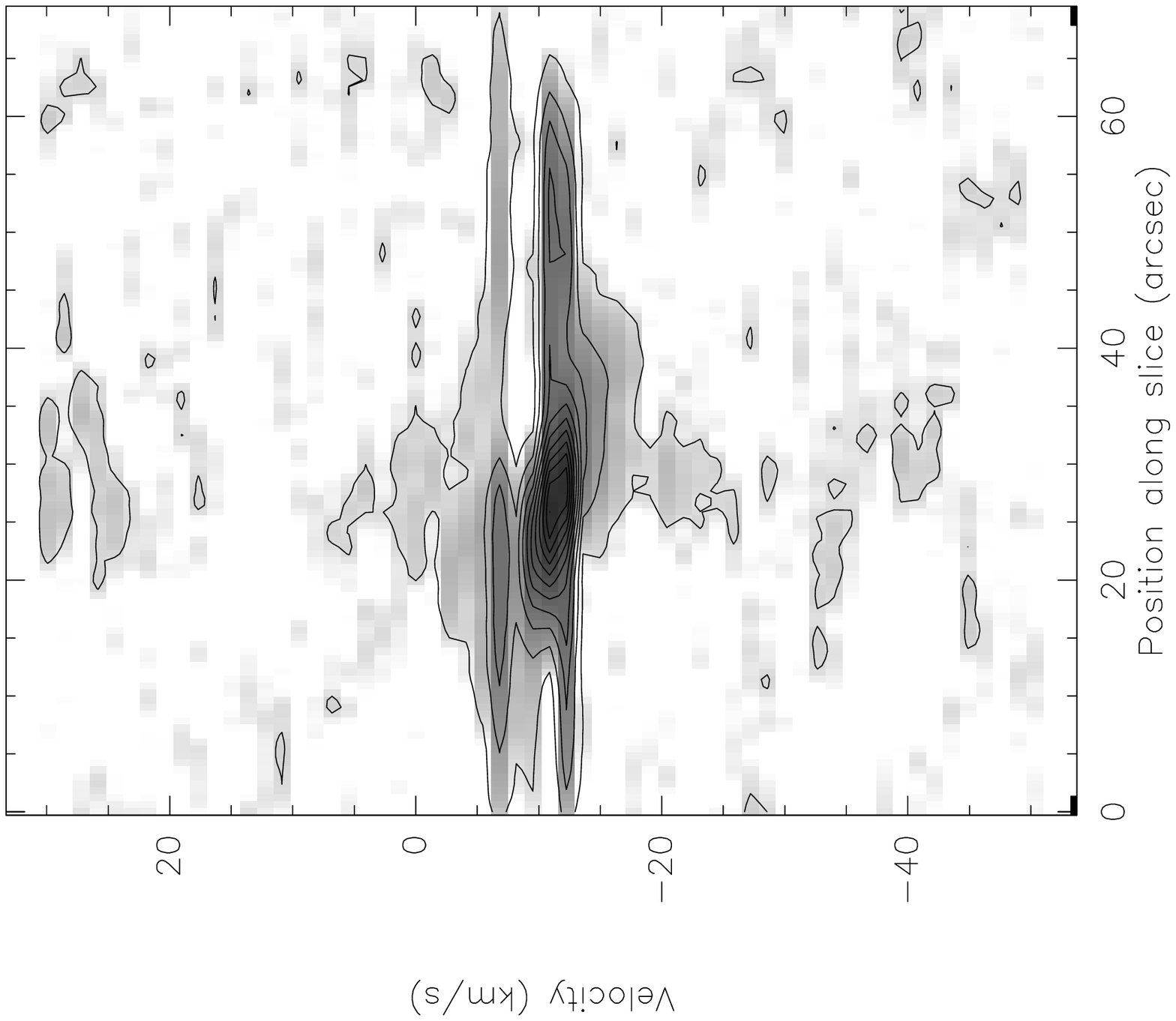]{Position velocity diagram in the $^{13}$CO 
J=1-0 transition along the axis of the Cep E outflow. The lowest contour 
correspond to a 2-$\sigma$ level.
\label{f10}}

\figcaption[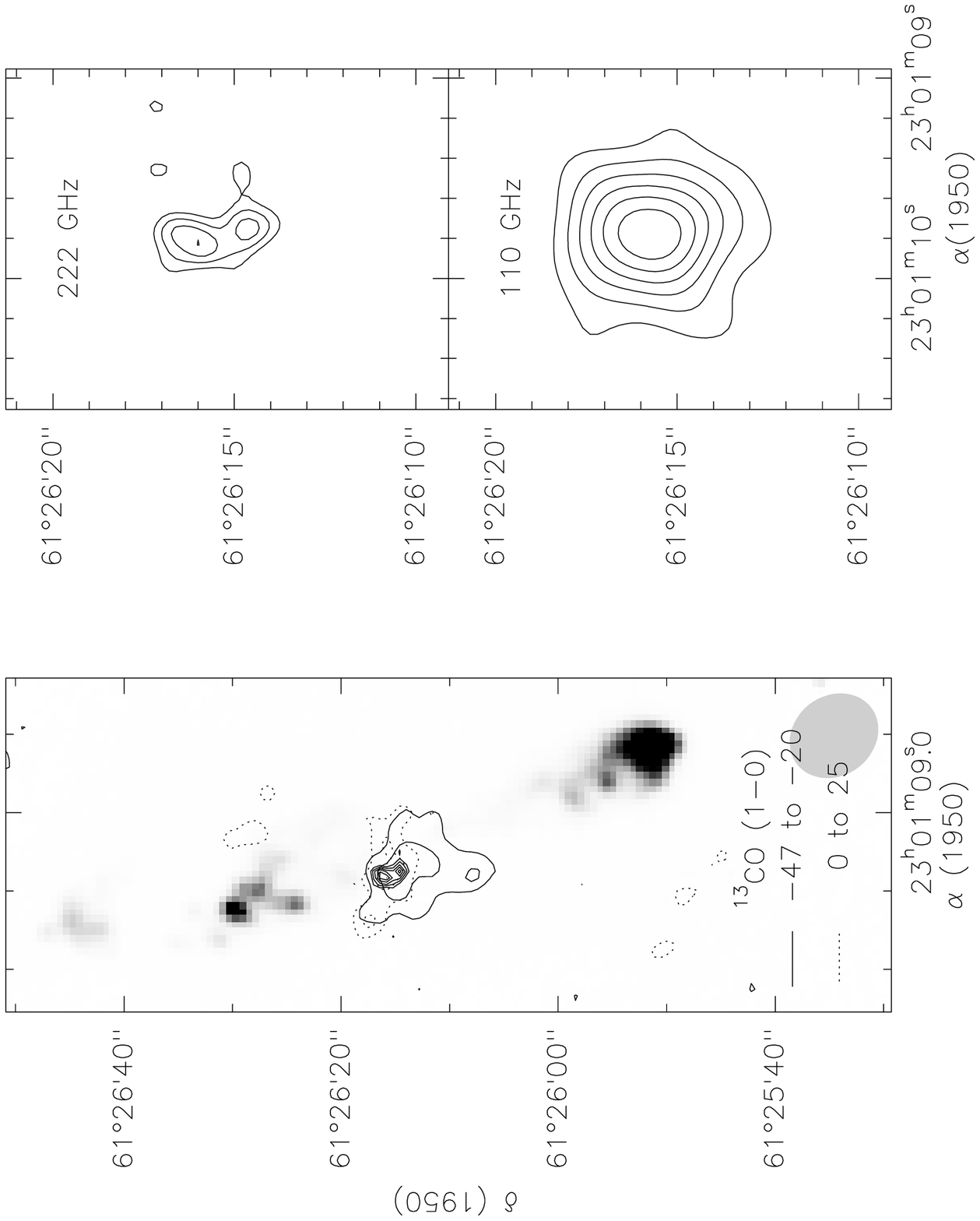]{Left: same as Figure 9 but for the high
velocity gas. The red lobe corresponds to the gas with velocities 
between 0 and 20 \kms~ (dotted contours)
and the blue lobe to gas with velocities between -47 and -20 \kms (solid
contours).
The contour levels are 10 and 15 in both cases. The detections are at 
5-7 $\sigma$ level. Right: 222 GHz and 110 GHz
continuum image centered on IRAS 23011+6126. The flux levels are 
from 0.02 to 0.05 by 0.01
and from 0.003 to 0.02 by 0.003 respectively (in Jy/beam).
\label{f11}}

\figcaption[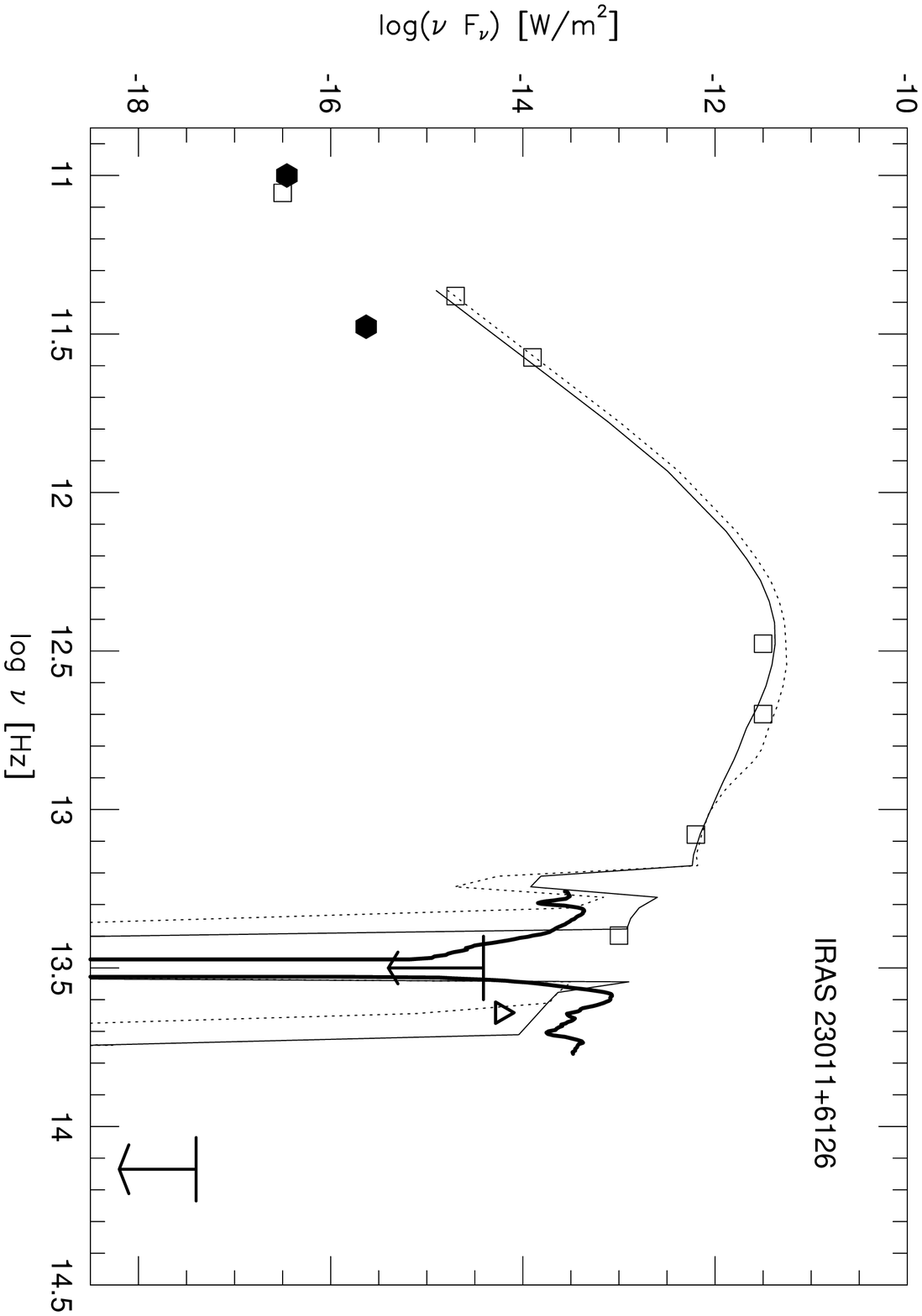]{Spectral energy distribution of 
IRAS 23011+6126. Data points are taken from Ladd \& Howe 1997 
(submillimeter; squares), 
from Noriega-Crespo et al.~1998 (IRAS and ISOCAM; open triangles and upper 
limit) and from our new OVRO results (filled symbols). The thick solid line 
represents  the CVF spectra extracted at the source position. A model 
assuming a dust opacity dominated by bare silicates, at a temperature of 18 K 
and n = 7.5$\times10^4$\cc~is shown in solid line. The dotted line is for 
silicates with a thin ice mantle at the 18 K and n = 6$\times 10^4$\cc.
\label{f12}} 

\figcaption[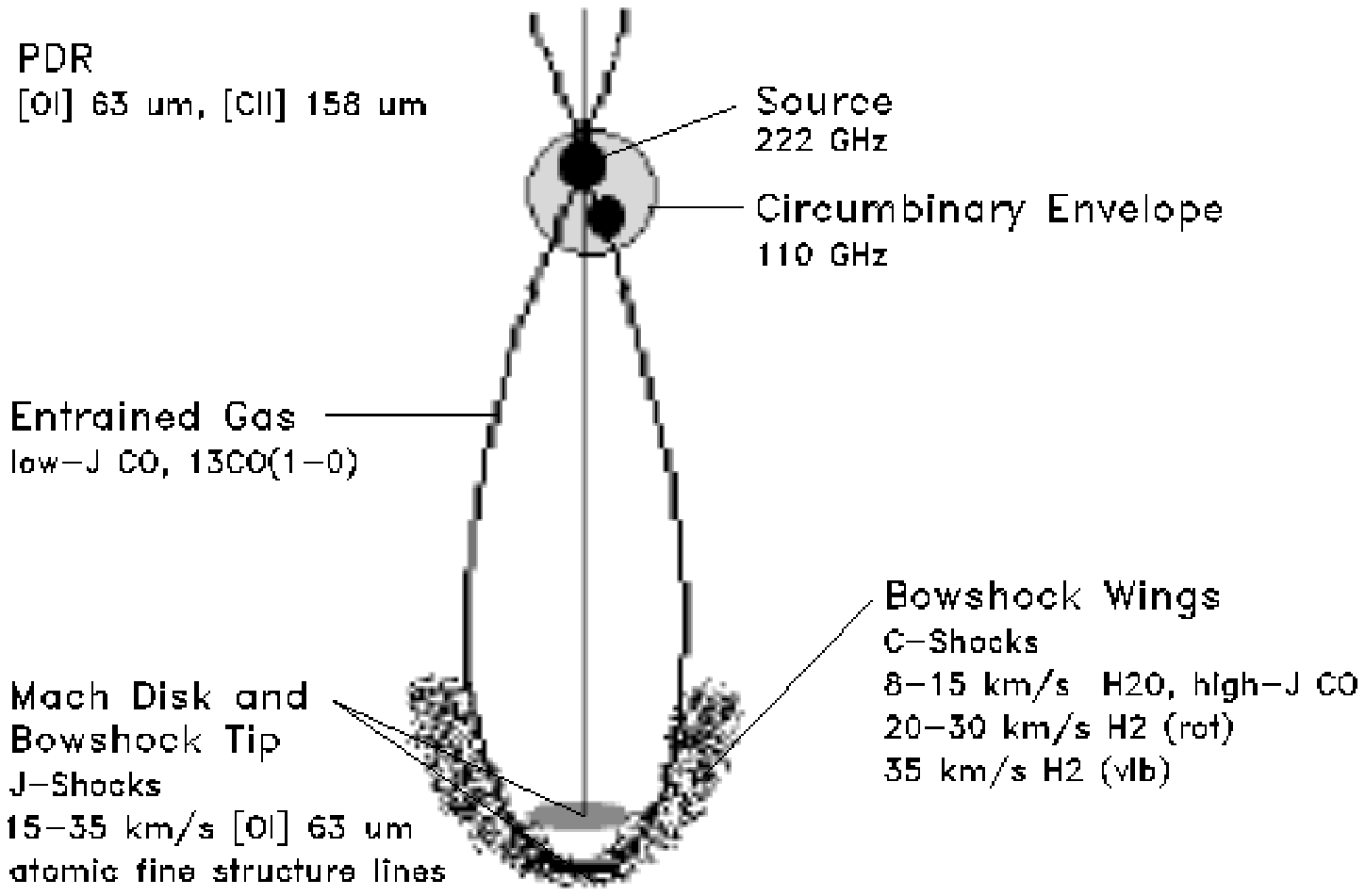]{Schematic view of the different emitting  
regions in the Cep E outflow adopting a jet-driven model.
\label{f13}} 

\clearpage

\begin{figure}[v]
\plotone{amm_fig01.ps}
\end{figure}
\clearpage

\begin{figure}[v]
\plotone{amm_fig02.ps}
\end{figure}
\clearpage

\begin{figure}[v]
\plotone{amm_fig03.ps}
\end{figure}
\clearpage

\begin{figure}[v]
\plotone{amm_fig04.ps}
\end{figure} 
\clearpage

\begin{figure}[v]
\plotone{amm_fig05.ps}
\end{figure}
\clearpage

\begin{figure}[v]
\plotone{amm_fig06.ps}
\end{figure}
\clearpage

\begin{figure}[v]
\plotone{amm_fig07.ps}
\end{figure}
\clearpage

\begin{figure}[v]
\plotone{amm_fig08.ps}
\end{figure}
\clearpage

\begin{figure}[v]
\plotone{amm_fig09.ps}
\end{figure}
\clearpage

\begin{figure}[v]
\plotone{amm_fig10.ps}
\end{figure}
\clearpage

\begin{figure}[v]
\plotone{amm_fig11.ps}
\end{figure}
\clearpage

\begin{figure}[v]
\plotone{amm_fig12.ps}
\end{figure}
\clearpage

\begin{figure}[v]
\plotone{amm_fig13.ps}
\end{figure}
\clearpage


\clearpage

\end{document}